\documentclass[%
 aip,
 jmp,%
 cha,%
 amsmath,amssymb,
preprint,%
]{revtex4-1}

\usepackage[english]{babel}

\usepackage{graphicx}
\usepackage{subfigure}
\usepackage{dcolumn}
\usepackage{bm}
\usepackage[mathlines]{lineno}

\usepackage{float}
\usepackage{bigstrut}

\usepackage{amsthm}

\theoremstyle{definition}
\newtheorem{defn}{Definition}[section]

\begin{document}


\title[]{ Hyperbolic Covariant Coherent Structures in two dimensional flows}

\author{Giovanni Conti}
 \email{giovanni.conti@uni-hamburg.de}
\author{Gualtiero Badin}%
 \email{gualtiero.badin@uni-hamburg.de}
\affiliation{ 
Institute of Oceanography, Center for Earth System Research and Sustainability (CEN), University of Hamburg, Hamburg, Germany
}%


\date{\today}

\begin{abstract}
A new method to describe hyperbolic patterns in  two dimensional flows is proposed. The  method is based on the Covariant Lyapunov Vectors (CLVs), which have the  properties to be covariant with the dynamics, and thus being mapped by the tangent linear operator into another CLVs basis, they are norm independent, invariant under time reversal and can be  not  orthonormal. CLVs can thus give a more detailed information on the expansion and contraction directions of the flow than the Lyapunov Vector bases, that are instead always orthogonal. 
We suggest a definition of  Hyperbolic Covariant Coherent Structures (HCCSs), that can be defined on the scalar field representing the angle between the CLVs.  HCCSs can be defined for every time instant and could be useful to understand the long term behaviour of particle tracers. 
 We consider three examples: a simple autonomous Hamiltonian system, as well as the non-autonomous ``double gyre'' and Bickley jet, to see how  well the angle is able to describe  particular patterns and barriers.  We compare the results from the HCCSs with other coherent patterns defined on finite time  by  the Finite Time Lyapunov Exponents (FTLEs), to see how the behaviour of these structures change asymptotically. 
  \end{abstract}

\maketitle

\begin{quotation}

\end{quotation}

\section{\label{sec1:1}Introduction}
In the paradigm of chaotic advection,  the trajectories of passive tracers can  be complex even when the velocity field of the flow is  simple. This is the case, for example, for time dependent two-dimensional flows or even steady three-dimensional flows, like the celebrated ABC flow \cite{dombre1986chaotic}. However, even flows with  complicated time dependent structure allow for the formation of coherent patterns that influence the evolution of  tracers. These structures are common in nature,  appearing both at short and long time scales as well as small and large spatial scales. Remarkable examples of these structures  are  eddies and jets in the ocean and atmosphere,  the Gulf Stream current, and ring clouds \cite{Haller2015}. These patterns can influence, for example, the evolution of nutrients as well as oil spills and other pollutants. Furthermore,  these coherent structures could act as local inhibitor for the energy transfer between scales \cite{Kelley2013}.  
They also appear  in other planets and other astrophysical systems,  such as the jets  on the surface of Jupiter, Saturn and other gaseous planets,  and  in the solar photospheric flows \cite{Chian2014}.   
These structures, often referred as Lagrangian Coherent Structures (LCSs), can shed light on the mixing and transport properties of a particular system on a finite time interval. The term ``Lagrangian''  in their name is motivated by  the fact that they evolve as material lines with the flow \cite{Haller2015}. 
A particular kind of LCS is called Hyperbolic LCS (HLCS), and can be seen as the locally most attracting or repelling material lines  that characterize the dynamical system over a finite time interval. The term hyperbolic is just an analogy with the stability of fixed points in dynamical systems, since usually one  wants to study non-autonomous systems for which entities such as fixed points or stable and unstable manifolds are not defined. Moreover these systems are studied for finite time.

Early attempts to detect HLCSs were based on  Finite Time Lyapunov Exponents (FTLEs), that measure the rate at which initial conditions (or, equivalently, tracer particles) separate locally after a given interval of time\cite{Pierrehumbert1993,Doerner1999,Haller2000,Haller2001,Lapeyre2002,Shadden2005,Lekien2007,Sulman201377}. One of the first rigorous definitions of the LCSs was based on ridges of the FTLEs \cite{Shadden2005}.
FTLEs and the so defined LCSs were used to study for example Lagrangian dynamics of atmospheric jets \cite{Rypina2007} and oceanic stirring by mesoscale eddies \cite{Rypina2010,Vera2008,Waugh2008,Bettencourt201273,Harrison2012,Waugh2012}, describing for example the chaotic advection emerging by mixed layer instabilities and its sensitivity to the vertical shear \cite{Mukiibi}. 
 A similar method to detect coherent structures  uses  the Finite Size Lyapunov Exponents (FSLEs), that represents the separation rate of particles given a specific final distance \cite{Joseph2002,d'Ovidio,Cencini2013}. However, several counter examples are available in which both FTLEs and FSLEs ridges  fail in characterizing the  LCSs \cite{Karrasch2013,Karrasch2015}. Although the FTLEs field remain a popular diagnostics of chaotic stirring, other methods now are available to detect LCSs, which include for example the so called Lagrangian descriptors
\cite{Mancho2013,lopesino2017theoretical}, which are  based on  integration along trajectories, for a finite time,  of an intrinsic bounded positive geometrical and/or physical property of the trajectories themselves. Notice however that the method of Lagrangian descriptors is not objective \cite{Vortmeyweyr}; the connection between the Perron-Frobenius operator and almost invariant coherent sets of non-autonomous dynamical systems defined over infinite times \cite{Froyland20101527}, the use of braids \cite{Thiffeault2010}, and the extrema of trajectories \cite{mundel2014new}.  Other Fast Indicators (FIs) besides FTLEs and FSLEs, i.e. computational diagnostics that characterize chaos quickly and can be used to determine coherent structures, are the Smaller (SALI) and Generalized (GALI) Alignment Indices \cite{SKOKOS200730}, the Mean Exponential Growth rate of Nearby Orbits (MEGNO) \cite{Cincotta2000}, and the Finite Time Rotation Number (FTRN) \cite{SZEZECH2013452, Szezech2012, Szezech2009}.

Particularly promising for the detection of coherent structures is a variational theory that considers the extremum properties of a specific repulsion rate function
\cite{Haller2011,Farazmand2012439,Karrasch20121470,Farazmand2012,Farazmand2013,Karrasch2015,Blazevski201446}. Using this theory it has been further shown  that HLCSs can be described using geodesics \cite{Haller2012}. In this way, HLCSs can be represented as minimizers of a material length, with specific boundary conditions for the variation function. Another geodesic theory describes  HLCSs in terms of shearless transport barrier that minimize the average shear functional \cite{Farazmand2013,Farazmand201444}.
These theories are based on the computation of shrink/stretch-lines (tensorlines), thus  trajectories along the eigenvectors of the deformation tensor also called  Cauchy Green Tensor (CGT). 
Geodesic theories are also able to detect other two kinds of LCSs called Parabolic and Elliptic,  which are  however of no  interest for the present work.

All these methods such as FTLEs, FSLEs, or the variational and geodesic theories, aim to find particular structures on the flow, among these the most repelling, or attracting, structures  in the flow on a finite time interval, the HLCS.  Particle tracers around these most influential material structures in finite time  are maximally repelled or attracted. However, changing the finite time interval under study also changes  the dynamical system and the correspondent structures emerging from the flow. Recent effort has been done in the understanding the  instantaneous  most influential coherent structures, Objective Eulerian Coherent Structures (OECSs), using a method that is not based on a  finite time interval of evolution \cite{MattiaOE,serra_haller_2017,Serra20160807}.
The tracer particles could be maximally repelled for a short time, but on long time they could have a different behaviour. 
Is it  natural to wonder what happen asymptotically to the  tracer particles? Is it possible to find some coherent structures that suggest the asymptotic behaviour of the tracer particles?

In this work we thus propose an  alternative method to detect coherent patterns emerging in chaotic advection, which is based on the Covariant Lyapunov Vectors (CLVs).  CLVs  were first introduced  by Oseledec \cite{Oseledec1968} and Ruelle \cite{Ruelle1979}, but for a long time they have received very little attention due to the lack of an efficient algorithm to compute them. Only in the last decade   the computation of such vectors has become possible \cite{Wolfe2007, Ginelli2007, Froyland201318}, and CLVs  have been used  to investigate e.g. the motion of  rigid disk systems \cite{Bosetti2010}, convection \cite{Xu2016}, and other atmospheric phenomena \cite{Wolfe2008,Schubert2015,Schubert2016,Vannitsem2016}. For other theoretical discussions or for reviews on CLVs, see e.g. Refs.
\cite{Hong2008,Hong2010,Kuptsov2012,Sala2012,Ginelli2013}.  Unlike the Lyapunov Exponents (LEs),  which are time independent, the correspondent vectors, also known as  forward and backward Lyapunov Vectors (LVs), do depend on time. The LVs are orthonormal and  their direction can thus give only limited information about the local structure of the attractor.
LVs also depend on the chosen norm, they are not invariant under time inversion and they are not covariant, where covariance is here defined as the property of  the forward (backward) LVs  to be mapped by the tangent dynamics in forward (backward) LVs at the image point \cite{Kuptsov2012}. Differently from the LVs,  CLVs  are norm independent, invariant for temporal inversion and covariant with the dynamics, making them thus mapped by the tangent linear operator into another CLVs bases during the evolution of the system.  They are not orthonormal and their directions can thus probe better the tangent structure of the system. 

All this intrinsic information can be summarized using the angle between the CLVs, a scalar field that allows to investigate the spatial structures of the system. The need to pay more attention to the directions between LVs, backward and forward, has also been suggested for the study of turbulence \cite{Lapeyre2002} and for the definition of a diagnostic quantity for the study of mixing, the Lyapunov's diffusion \cite{dOvidio2009}. Using three simple examples, we show that the attracting and repelling barriers tend to align along the paths on which the  CLVs  are orthogonal. 
The directions of the CLVs along these maxima provide thus information on the attracting or repelling nature of the barriers and can be related to the geometry of the system. Using the CLVs we can define structures, at a given time instant, that are asymptotically the most attractive or repelling. Furthermore, since these structures can be defined for every time instant, it is possible to follow the formation of  coherent structures during the evolution of the flow.

In section \ref{sec2:1} we discuss the theory behind the CLVs, and suggest a definition for coherent structures that give asymptotic information.  The strategy will be to make use of the scalar quantity defined by the angle between the CLVs to locally identify the structures that asymptotically are maximally attractive or repulsive. In section \ref{sec3:1}  we use the CLVs    to identify particular patterns in three different systems and we compare the results with FTLEs fields. Finally, in section \ref{sec4:1} we summarize the conclusions.

\section{\label{sec2:1}Covariant Lyapunov Vectors}

In this section we summarize the theory behind CLVs for two dimensional flows. For a more general and detailed review see for example Ref. \cite{Kuptsov2012}.
Let the open set $D\subset\mathbb{R}^2$ be the domain of the flow,  $t\in\mathbb{R}$ the time and  ${\bf v}({\bf x}, t)$ a vector velocity field in $D$. The dynamical system that describes the motion of a tracer advected by the flow is thus
\begin{subequations}
\begin{eqnarray}
&&\frac{d{\bf x}({\bf x}_0,t_0;t)}{dt}= {\bf v}({\bf x}({\bf x}_0 ,t_0; t),t),\\
&&{\bf x}({\bf x}_0 ,t_0; t_0)={\bf x}_0,
\end{eqnarray}
\label{ds}
\end{subequations}
where ${\bf x}({\bf x}_0 ,t_0,t)\in D$ is the trajectory of the tracer starting at the point ${\bf x}_0$ at time $t_0$.
To \eqref{ds} is associated the flow map $\boldsymbol\phi_{t_0}^{t}({\bf x}_0)$
\begin{equation}
\begin{split}
&\boldsymbol\phi_{t_0}^{t}: D\to D,\\
&{\bf x}_0 \to {\bf x}({\bf x}_0,t_0,t),
\end{split}
\label{fm}
\end{equation}
that maps the initial position ${\bf x}_0$ at time $t_0$ to the position ${\bf x}({\bf x}_0,t_0,t)$ at time $t$. 
It should be noted that the dependence on the initial condition  is very important here, since the vectors will be considered as function of  time and of the initial positions.  In the following, the contracted form ${\bf x}={\bf x}({\bf x}_0 ,t_0,t)$ will be used.

At each point ${\bf x}\in D$ we can identify the tangent space $T_{{\bf x}}D\subset\mathbb{R}^2$. Infinitesimal perturbations, ${\bf u}(t) \in T_{{\bf x}}D $, to a trajectory of this system can be described by the linearized system
\begin{subequations}
\begin{eqnarray}
&&\frac{d{\bf u}(t)}{dt}= {\bf J}(t)\,{\bf u}(t),\\
&&{\bf u}(t_0)={\bf u}_0,
\end{eqnarray}
\label{td}
\end{subequations}
where ${\bf J}(t)\in \mathbb{R}^{2\times2}$ is the Jacobian matrix composed by the derivatives of the vector field ${\bf v}({\bf x}, t)$ with respect to the component of the vector ${\bf x}$. Using the fundamental matrix ${\bf M}(t)\in \mathbb{R}^{2\times2}$, of  \eqref{td}, that satisfies 
\begin{subequations}
\begin{eqnarray}
\frac{d{\bf M}(t)}{dt}&=&{\bf J}(t){\bf M}(t),\\
{\bf M}(0)&=&{\bf I}
\end{eqnarray}
\end{subequations}
we define the so called tangent linear propagator   
\begin{equation}
{\bf F}(t_0, t)={\bf M}(t){\bf M}(t_0)^{-1}.
\label{pdef}
\end{equation}
${\bf F}(t_0, t)$ maps a vector in ${\bf x}_0$ at time $t_0$ into a vector in ${\bf x}$ at time $t$ along the same trajectory of the starting system  \eqref{ds}, that is
\begin{equation}
{\bf u}(t)={\bf F}(t_0, t)\,{\bf u}(t_0).
\label{pd}
\end{equation}
According to \eqref{pdef}, the propagator is always nonsingular.
In terms of the flow map, the tangent linear propagator is
\begin{equation}
{\bf F}(t_0, t)=\nabla \boldsymbol\phi_{t_0}^{t}.
\label{pflow}
\end{equation} 

Exploiting Oseledec's Theorem \cite{Oseledec1968,Eckmann1985}, it is possible to characterize the system using quantities that are independent on $t$ or $t_0$.
By virtue of this theorem the far-future operator
\begin{equation}
{\bf O}^+(t_0)=\lim_{t\to+\infty} \left( {\bf F}(t_0,t)^\top{\bf F}(t_0, t) \right)^{1/2(t-t_0)}
\label{Wp}
\end{equation}
and the far-past operator
\begin{equation}
{\bf O}^-(t)=\lim_{t_0\to-\infty} \left( {\bf F}(t_0,t)^{-\top}{\bf F}(t_0, t)^{-1} \right)^{1/2(t-t_0)}
\label{Wm}
\end{equation}
are well defined quantities.
Note  that the product ${\bf F}^\top(t_0, t){\bf F}(t_0, t)$ determines the Euclidean norm of the tangent vectors in the forward-time dynamics (a similar role it is played by ${\bf F}^{-\top}(t_0, t){\bf F}^{-1}(t_0, t)$ for the backward-time dynamics), in fact,
\begin{equation}
\lvert\lvert {\bf u}(t)\rvert\rvert=\left[{\bf u}(t_0)^\top\left({\bf F}^\top(t_0, t){\bf F}(t_0, t)\right){\bf u}(t_0)\right]^{1/2}.
\end{equation}
Operators \eqref{Wp} and \eqref{Wm} probe respectively the future and past dynamics of a certain point, and share the same eigenvalues 
\begin{equation}
 \lambda_1 \ge \lambda_2, 
 \label{new lambda}
\end{equation} 
 that,  assuming ergodicity, are independent on time and space.  Each eigenvalue has multiplicity $m_i$ ($m_1+m_2=2$). Their  logarithms  correspond to the LEs of the dynamical system \eqref{ds}.  If the limits  in \eqref{Wp} and \eqref{Wm} are not considered, the resulting eigenvalues are time and space dependent  and are called FTLEs.

The two operators \eqref{Wp} and \eqref{Wm} can be evaluated at the same point in space at a given time $t$. The correspondent  eigenvectors, $\{ {\bf l}^+_1(t),\,\,  {\bf l}^+_2(t)\}$,  $\{{\bf l}^-_1(t),\,\,  {\bf l}^-_2(t)\}$ will define thus the forward and backward Lyapunov basis computed at the same time. 
 Conversely to the respective eigenvalues those bases are time dependent, depend on the chosen scalar product  and are not invariant under time reversal. Furthermore, these vectors are always orthogonal and give thus limited information on the spatial structure of the configuration space.
 
To overcome these issues, one can build particular spaces, the backward and forward Oseledec subspaces, defined as \cite{Oseledec1968}
\begin{equation}
L^-_i(t)=\text{span}\{{\bf l}^-_j (t)\lvert j=1,i\},\quad i=1,2, \quad L^-_{0}(t)=\emptyset, 
\label{lm}
\end{equation}
and
\begin{equation}
L^+_i(t)=\text{span}\{{\bf l}^+_j (t)\lvert j=i,2\},\quad i=1,2, \quad L^+_{0}(t)=\emptyset
\label{lp}
\end{equation}
In the forward dynamics, the generic vector ${\bf l}^+_i (t)$ grows or decays exponentially with an average rate $\lambda_i$.
If the system is  evolved forward in time, by means of the tangent linear propagator the evolution of the vector ${\bf l}^+_1 (t)$ will have a non-zero projection inside the space generated by   ${\bf l}^+_1(t')$, but it will also have a non-zero projection onto the space generated by  ${\bf l}^+_2 (t')$. On the other hand, ${\bf l}^+_2 (t)$ will be  transported onto the space generated by ${\bf l}^+_2 (t')$ and will have zero projection onto the space generated by ${\bf l}^+_1(t')$. Repeating similar arguments for the backward Lyapunov Vectors leads to the observation that
$L^-_i$ and $L^+_i$ are covariant subspaces,
\begin{subequations}
\begin{eqnarray}
L^-_i(t')  &=& {\bf F}(t,t') L^-_i(t),\\
L^+_i(t') &=& {\bf F}(t,t')L^+_i(t).
\end{eqnarray}
\end{subequations}
Vectors that are  covariant with the dynamics and invariant with respect to time reversal will be found at the intersection of the Oseledec subspaces, i.e. at
\begin{equation}
{\bf w}_i(t)= L^-_i(t)  \cap   L^+_i(t). 
\label{wo}
\end{equation}
These spaces, often referred as Oseledec splitting \cite{Ruelle1979, Oseledec1968, Ginelli2007, Ginelli2013}, are not empty \cite{Kuptsov2012}, and
their vectors, also called covariant Lyapunov vectors (CLVs), are  covariant with the dynamics, i.e.
\begin{equation}
{\bf F}(t, t'){\bf w}_i(t)={\bf w}_i(t').
\end{equation}
It should be noted that CLVs posses thus the properties of a semi-group.
These vectors have an asymptotic grow or decay with an average rate $\lambda_i$, so that their asymptotic behaviour can be summarized as
\begin{equation}
\lVert {\bf F}(t,t\pm \tau) {\bf w}_i(t) \rVert \approx e^{\pm \lambda_i \tau},
\label{asyCLVs}
\end{equation}
where $\tau=\lvert t'-t \rvert$, which shows their invariance under time reversal. 
Note also that these vectors do not depend on any particular norm. 


Equations  \eqref{lm}, \eqref{lp} and \eqref{wo} imply a simple relation between CLVs and LVs
in a two dimensional system, 
\begin{equation}
{\bf w}_1 \equiv {\bf l}^-_1\quad \text{and} \quad {\bf w}_2 \equiv{\bf l}^+_2,
\label{relV}
\end{equation}
the first CLV corresponds to the first backward Lyapunov vector, and the second CLV with the second forward Lyapunov vector. 

Furthermore, if the Jacobian matrix appearing in \eqref{td} is constant,  the CLVs are not just covariant, but  invariant with the dynamics. In this particular case in fact they correspond to the eigenvectors of the Jacobian and the eigenvalues of this matrix coincide with the LEs.

Since the CLVs highlight particular expansion and contraction directions at each point of the coordinate space, and these directions are not necessarily orthogonal,  they can be used to understand the geometrical structure of the tangent space. This geometric information can be summarized by  the scalar field of the angle $\theta(t)$ between the CLVs. Because the CLVs identify asymptotically the expansion and contraction directions sets of the tangent space associated at every point of the domain, the $\theta$ field represents a measure of the hyperbolicity of the system, that is a measure of the orthogonality between these two directions.
 Note that  the orientation of the CLVs is defined as arbitrary. The angle between the CLVs, 
 \begin{equation}
 \theta(t) \in \left[0,\,\,\frac{\pi}{2}\right],
 \end{equation}
is thus
\begin{equation}
\theta(t)=\cos^{-1}\left(\lvert{\bf w}_1(t)\cdot{\bf w}_2(t)\lvert\right), 
\end{equation}
where $\{{\bf w}_1(t),\,{\bf w}_2(t)\}$ are the first and the second CLVs \cite{Ginelli2007}.
It is interesting to point out that when $\theta(t)=\pi/2$, the CLVs reduce to LVs in two dimensions, and the backward and forward Lyapunov basis coincide.
If the computation of the angle is done at every point of the domain, and if we consider  the system  \eqref{ds} in which the initial conditions are varied in such a way that all the domain is spanned, we can build a field of the orthogonality between the expansion and contraction directions of the system.

In the following section we will show how CLVs, with their capability of probing the geometric structure of the tangent space, can be used to determine coherent structures that give asymptotic information on the tracer and then, how the CLVs highlights the mixing template of the flow. 

\subsection{\label{sec2:2} Hyperbolic Covariant Coherent Structures}

Several frameworks are available to study passive scalar mixing. Although all these methods aim to describe the mechanism underlying the chaotic advection, they have significantly different approaches. However, most of them share a fundamental feature called objectivity.
Objectivity is a fundamental requirement to define structures emerging from a flow.  In practical applications, objectivity can be used e.g. to move the reference frame into the reference frame of a coherent structure. In particular a structure can be considered objective if it is invariant under a coordinate changes of the form
\begin{equation}
{\bf \hat{x} }(t)={\bf Q}(t){\bf x}(t) + {\bf P}(t), 
\label{trans}
\end{equation}
where ${\bf Q}$ denotes a time-dependent orthogonal matrix and ${\bf P}$ a time-dependent translation \cite{Haller2015}.
The FTLEs, FSLEs and geodesic theory  define objective quantities. Sala et all. \cite{Sala2012} have shown that, generally, also for a linear transformation of coordinates the new angle non-linearly depends  on the  angle of the old reference system. However, for the particular class of transformation  we are interested in, (\ref{trans}), the angle $\theta$ is an invariant. This means that structures highlighted by $\theta$ are objective.

\begin{figure}
\centering
  \includegraphics{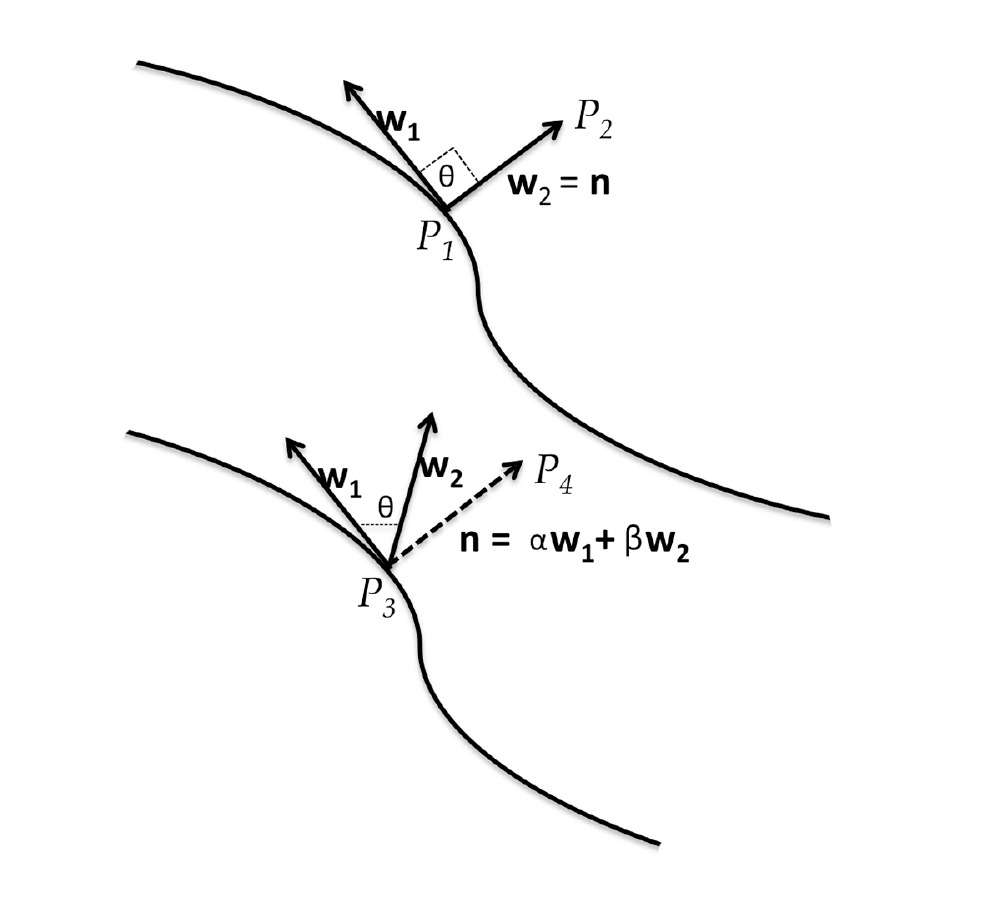}
\caption{Orientation of CLVs on a most attracting material line (upper line) and on a neighbouring material line (lower line). The distance $\lVert P_2-P_1\rVert$ decreases faster than $\lVert P_4-P_3\rVert$ since in the second curve the normal vector connecting the points $P_3$ and $P_4$ is expressed as a linear combination of both the CLVs.}
 \label{fig:1}
\end{figure}

To study the asimptotically most attractive or repelling behaviour of tracer particles  near a particular structure identified at a given instant of time, one can consider the lines ${\bf r}(s,t)$, with $s$ representing  the length parameter, defined by
\begin{subequations}
\begin{eqnarray}
{\bf r}_1'(s,t)&=&{\bf w}_1({\bf r}(s,t),t)\label{tlinea},\\
{\bf r}_2'(s,t)&=&{\bf w}_2({\bf r}(s,t),t),
\label{tlineb}
\end{eqnarray}
\label{tlineclv}
\end{subequations}
where the primes indicates the derivative with respect to $s$, and characterized by $\theta=\pi/2$ along their paths. In fact, in this case
\begin{equation}
\theta(t)=\frac{\pi}{2} \Rightarrow \quad  {\bf w}_1\parallel {\bf l}_1\quad  \text{and} \quad {\bf w}_2 \parallel {\bf l}_2,
\label{contens}
\end{equation}
where, ${\bf l}_1$ and ${\bf l}_2$ are Lyapunov vectors. In these circumstances the backward and the forward basis are coincident. A sphere of initial conditions around a point of the path will be deformed in an ellipsoid whose axis are aligned with these Lyapunov vectors.
If \eqref{tlineclv} are aligned along a ridge of the hyperbolicity field $\theta$, characterized by $\theta=\pi / 2$, then they are also pointwise the most attractive or repelling lines in terms of the  asymptotic behaviour of tracer particles. Consider in fact line (\ref{tlinea}) and its tangent CLV ${\bf w}_1$ as depicted in Figure \ref{fig:1}. 
Along this curve, for any point $P_1$ we can chose a point $P_2$ which is on the normal vector to the curve in $P_1$, ${\bf n}$. Because of \eqref{new lambda}, the distance between these two points decreases as
\begin{equation}
\lVert P_2-P_1\rVert \propto  \lVert {\bf w}_2 \rVert \approx e^{ \lambda_2 (t'-t)}.
\end{equation}
Consider  now a point $P_3$ on a nearby curve, and a point $P_4$ laying on the normal vector to the curve in $P_3$. The normal vector ${\bf n}$ at $P_3$ can be written as a linear combination of ${\bf w}_1$ and ${\bf w}_2$ and, considering again \eqref{new lambda}, one has thus
\begin{equation}
\lVert P_4-P_3\rVert =  \lVert \alpha{\bf w}_1 +  \beta{\bf w}_2 \rVert \approx e^{ \lambda_1 (t'-t)} > e^{ \lambda_2 (t'-t)}.
\end{equation}
The previous computation can be repeated analogously for ${\bf w}_2$.
The same reasoning holds for ${\bf w}_2$.

 It must be remarked that  (\ref{tlineb}) gives asymptotic information, but is defined for a particular instant of time.   For every time instant  the lines \eqref{tlineclv}, together with the orthogonality condition $\theta(t)=\pi / 2$, describe the so called tensorlines, and can be seen as the asymptotic version of  shearless barriers \cite{Farazmand201444}.

Taking into account these properties of the CLVs, we propose a  definition to describe these asymptotic coherent patterns:
\begin{defn}[Hyperbolic Covariant Coherent Structures (HCCSs)]  At each time $t$, a \textbf {Hyperbolic Covariant Coherent Structure}, is an isoline  of the hyperbolicity scalar field $\theta$ at the level $\theta=\pi/2$. Its attractive or repelling nature  is determined by the CLVs aligned with it.  \label{defHCCS}
 \end{defn}

It is important to emphasize the difference between the information provided by FTLEs and the one provided by CLVs. FTLEs are the finite time version of Lyapunov's exponents and, as shown in the appendix, are calculated as a mean time of the logarithms of the separation of two trajectories that start from near points. The FTLEs therefore give an average information over the time interval considered. Moreover, if the time interval is long enough, and the analyzed region can be considered ergodic, the dependence on initial conditions is lost and therefore the possibility of displaying possible structures.
CLVs, on the contrary, as well as LVs, depend on a time instant and do not converge to the same value. CLVs therefore allow the definition of  instant structures, which, however, give asymptotic information. 

This approach presents some numerical challenges, CLVs are not fast to be computed as the FTLEs, and numerically, in order to identify the isolines, it is necessary to consider an interval of values for the angle. The HCCSs are thus computed here considering isoregion of $\theta\in[\frac{\pi}{2}-\delta,\,\,\frac{\pi}{2}]$, where $\delta$ is $0.087$ rads.

\section{\label{sec3:1}Numerical Examples}

The algorithm used in this work is the same presented in  Ref. \cite{Ginelli2007}, so only a schematic of its structure will be presented here. This algorithm can compute a large number of CLVs converging exponentially fast when  invertible dynamics is considered.
The basic idea is that, if the backward Lyapunov vectors basis $\{ {\bf l}^-_1(t),\,\, {\bf l}^-_2(t)\}$ is evolved by means the the operator ${\bf F}$, it is always possible to keep it orthonormalized with a $QR$ decomposition and store the corresponding upper triangular projection matrix.  It is then possible to exploit  the informations contained in these upper triangular matrices  to evolve backward in time arbitrary vectors  that will converge to the CLVs \cite{Ginelli2007}. This is done in five different phases, which are described in Appendix \ref{sec3:1-0}. 

In the next we investigate  the hyperbolicity field $\theta$ for three different examples, which include one autonomous Hamiltonian flow and two non-autonomous two dimensional flows. The HCCSs  emerging from the angle between the CLVs are compared with the FTLEs field, while their 
attractive or repelling nature  will  be discussed in terms of the CLVs. The algorithm for the search of the CLVs and the computation of the FTLEs explore the same dynamics, because it is used the same time interval. However,  the final CLVs are found  just for a temporal subinterval of the whole time used, see Appendix \ref{sec3:1-0} and  Appendix \ref{sec3:1-1}.

 Technical details about the three different examples are included in the Appendix \ref{sec3:1-1} for a clearer description of the physical results in the following.


\subsection{\label{sec3:2-1}A simple autonomous Hamiltonian system}
\begin{figure}
\centering
 \includegraphics[scale=1]{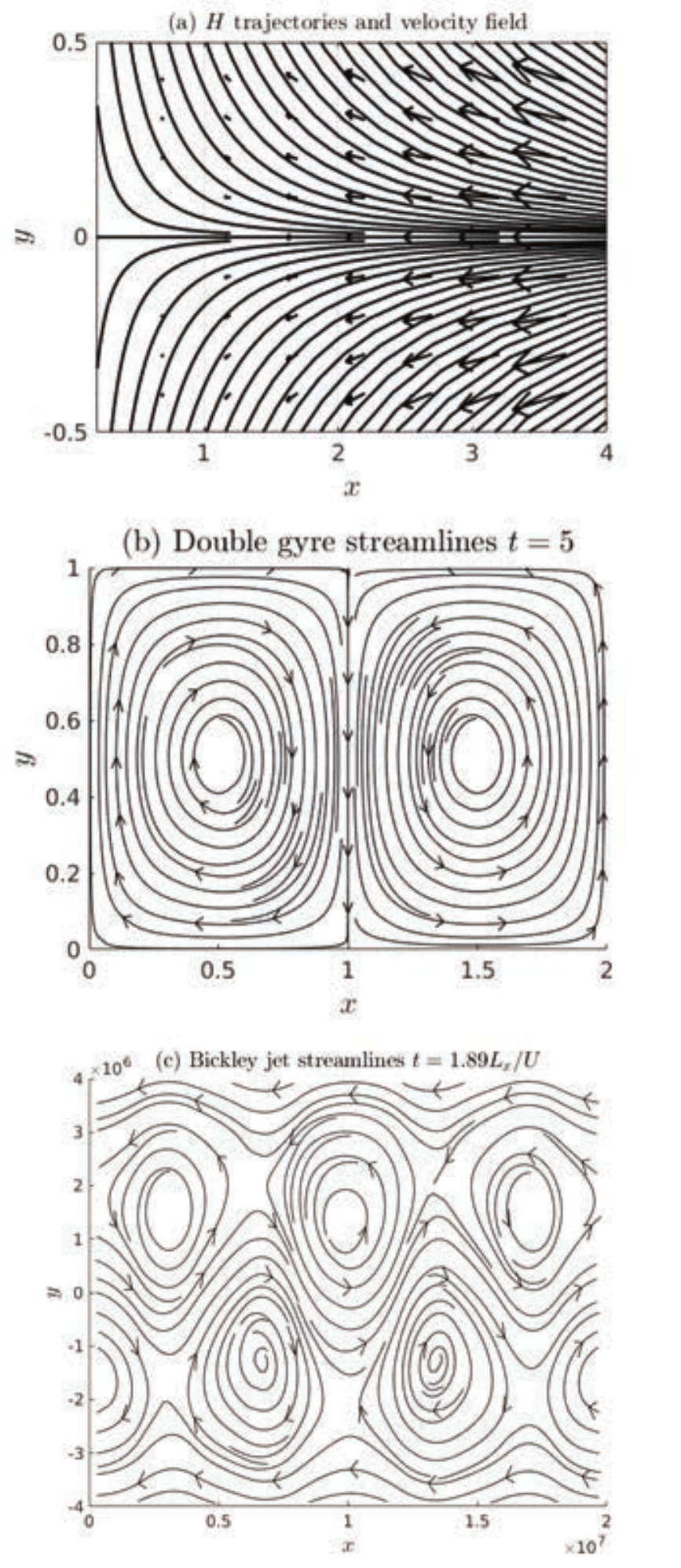}
\caption{Panel (a): trajectories of the Hamiltonian system  \eqref{Htraj} (solid line).  The trajectory $y=0$ represents a repulsive barrier for the system.  Notice that the arrows indicate the flow velocity field,  which is expressed on the right hand side of \eqref{Heq}, and not the CLVs that are shown separately. Panels (b) and (c): streamlines for the double gyre and the Bickley jet, respectively.}
 \label{fig:2}
\end{figure}

In this preliminary example we investigate the HCCSs in  an autonomous Hamiltonian flow map. 
The time-independent Hamiltonian, corresponding to the streamfunction of the flow, is 
\begin{equation}
H=\frac{x^2 y}{2},
\label{H}
\end{equation}
so that the dynamical  equations of motion are 
\begin{subequations}
\begin{eqnarray}
\frac{d x}{dt}&=&-\frac{x^2}{2},\\
\frac{d y}{dt}&=&xy,
\end{eqnarray}
\label{Heq}
\end{subequations}
with $x(t_0)=x_0$, and $y(t_0)=y_0$.
Integration of Eq. \eqref{Heq} yields the trajectories of the system, shown in Figure \ref{fig:2}a, with flow map 
\begin{equation}
\boldsymbol\phi_{t_0}^t(x_0,y_0) =\begin{bmatrix}
x(t)=-\frac{2x_0}{2+x_0 t}\\
y(t)=y_0\left(1+\frac{x_0 t}{2}\right)^2
                                    \end{bmatrix}.
\label{Htraj}
\end{equation}
Only the positive $x>0$ axis is considered, as the solution on the negative part blows up in finite time. As Figure \ref{fig:2}a shows, the $y=0$ axis  clearly divides the dynamics of the system into two different regions. There is no flux  across this line, and the positive and negative $y$ regions are completely separated at each time. Furthermore,  $y=0$ is the only material repelling line. Notice that in the Figure, the arrows do not indicate the CLVs but are just indicators of the magnitude of the flow.
The trajectory $y=0$ behaves thus exactly as a material barrier dividing the system in two different regions, remains coherent at all time, and repels every close trajectories. 
We  apply the CLVs theory to see if the angle between that vectors is able to detect such kind of structure.

\begin{figure}
\centering
  \includegraphics[scale=0.9]{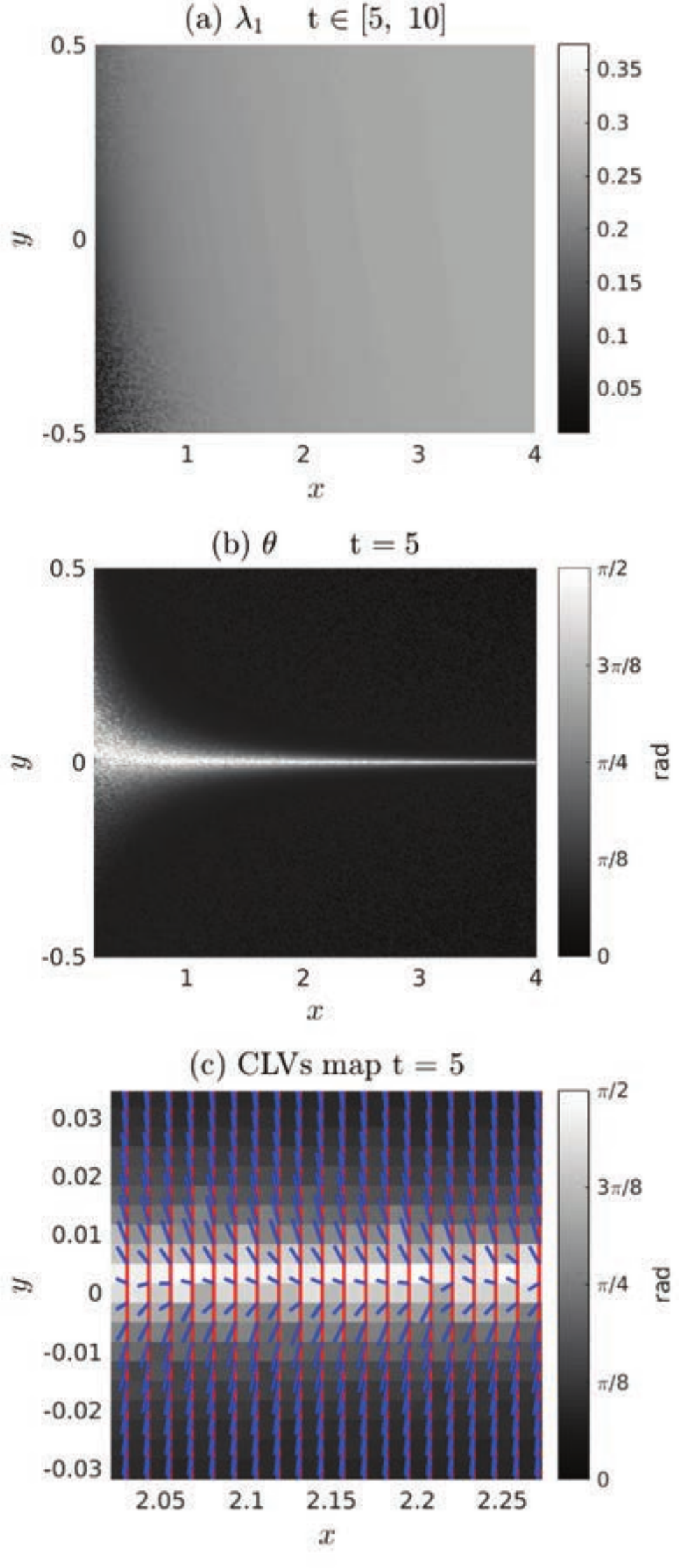}
\caption{FTLEs and  CLVs fields for the Hamiltonian system \eqref{Heq}. (a) Maximum FTLEs field computed from \eqref{avel}; (b) $\theta$ field; (c) zoom of the domain to show the CLVs fields.  Arrows are not used due to the arbitrariness of the orientation of the CLVs. The red lines are associated to ${\bf w}_1$, the expansion direction, while the blue lines are associated to with ${\bf w}_2$, the contraction direction. }
\label{fig:3}
\end{figure}

Figure \ref{fig:3}a shows the maximum FTLEs field computed in the time interval $t=[5,\,\,10]$ using definition \eqref{avel}. This field, referred to the initial grid conditions,  does not highlight  the $y=0$ barrier. Notice that if a bigger domain in the $y$ direction would have been selected, the FTLEs field would have showed a large scale modulation with larger values aligned along $y=0$. Figure \ref{fig:3}b, shows the distribution of the CLVs angles at $t=5$.  It is important to stress the fact that the FTLEs depend on a finite time interval while CLVs, and then $\theta$, on a particular instant of the time. The  hyperbolicity field clearly shows values close to $\pi/2$ along the $y=0$ axis. The direction of the first (red) and second (blue) CLV is shown for an enlargement of a region close to this structure in Figure \ref{fig:3}c. It is visible that the second CLV, that characterize the contraction direction, aside for numerical fluctuations, are aligned with the $\theta=\pi/2$ structure, indicating a contraction direction and thus a repelling structure, i.e. a barrier, forward in time.
Notice also that near $x=0$, the velocity field  is close to zero, so that the barrier there is not well defined. This suggests that the dynamics near the origin evolves with a different time scale. 

Along the $x$ axis, $\theta=\pi/2$  remains constant for all the time. Consider in fact the propagator \eqref{pflow}
\begin{equation}
{\bf F}(t, t')=\begin{bmatrix}
1/B^2 & 0\\
B t' y_0 & B^2
                                    \end{bmatrix},
\label{FH}
\end{equation}
where 
\begin{equation}
B=1+t' x_0/2.
\end{equation}
Along $y=0$, both the tangent linear propagator 
\begin{equation}
{\bf F}(t, t')=\begin{bmatrix}
1/B^2 & 0\\
0 & B^2
                                    \end{bmatrix}.
\label{FHD}
\end{equation}
and the CGT \eqref{CGT}
\begin{equation}
\begin{split}
{\bf C}(t, t')
                    &=\begin{bmatrix}
                    1/B^4 & 0\\
                     0 & B^4
                                    \end{bmatrix}, \quad y_0=0,
\end{split}
\end{equation}
are diagonal. If at time $t$ one has $\theta(t)=\pi/2$ for $y=0$, then 
\begin{equation}
{\bf w}_1(t)=\boldsymbol\xi_1=\begin{bmatrix}
0\\
1
                                     \end{bmatrix}\,\, \text{and} \,\,\,\,
                                     {\bf w}_2(t)= \boldsymbol\xi_2=\begin{bmatrix}
1\\
0
                                     \end{bmatrix},
\end{equation}
 where $\boldsymbol\xi_1$ and $\boldsymbol\xi_2$ are the eigenvectors of the CGT, and at time $t'>t$ 
\begin{equation}
\begin{split}
\cos(\theta(t')) & \propto {\bf F}(t, t'){\bf w}_1(t)\cdot{\bf F}(t, t'){\bf w}_2(t)\\
                      &= {\bf w}_1(t)\cdot{\bf C}(t, t'){\bf w}_2(t)\\
                      &=0.
\end{split}
\end{equation}
This shows that $y=0$ is a repelling HCCSs.

\subsection{\label{sec3:2-2}Double gyre}

The previous analysis is now applied to a time dependent system  called double gyre flow \cite{Shadden2005,Onu201526}, which is  a simplification of observed geophysical flows consisting of  two counter rotating vortices that expand and contract periodically. The tracers moving in this flow satisfy the following dynamical  equations
\begin{subequations}
\begin{eqnarray}
&&\frac{d x}{dt}=-\pi A \sin(\pi f(x,t))\cos(\pi y),\\
&&\frac{d y}{dt}=\pi A \cos(\pi f(x,t))\sin(\pi y)\frac{df(x,t)}{dx},\\
&&f(x,t)=\epsilon\sin(\omega t)x^2+(1-2\epsilon\sin(\omega t))x,
\end{eqnarray}
\label{DV}
\end{subequations}
with initial condition $x(0)=x_0$, $y(0)=y_0$. 
The streamline of the flow at $t=5$ are shown in Figure \ref{fig:2}b.

\begin{figure}
\centering
  \includegraphics[scale=1]{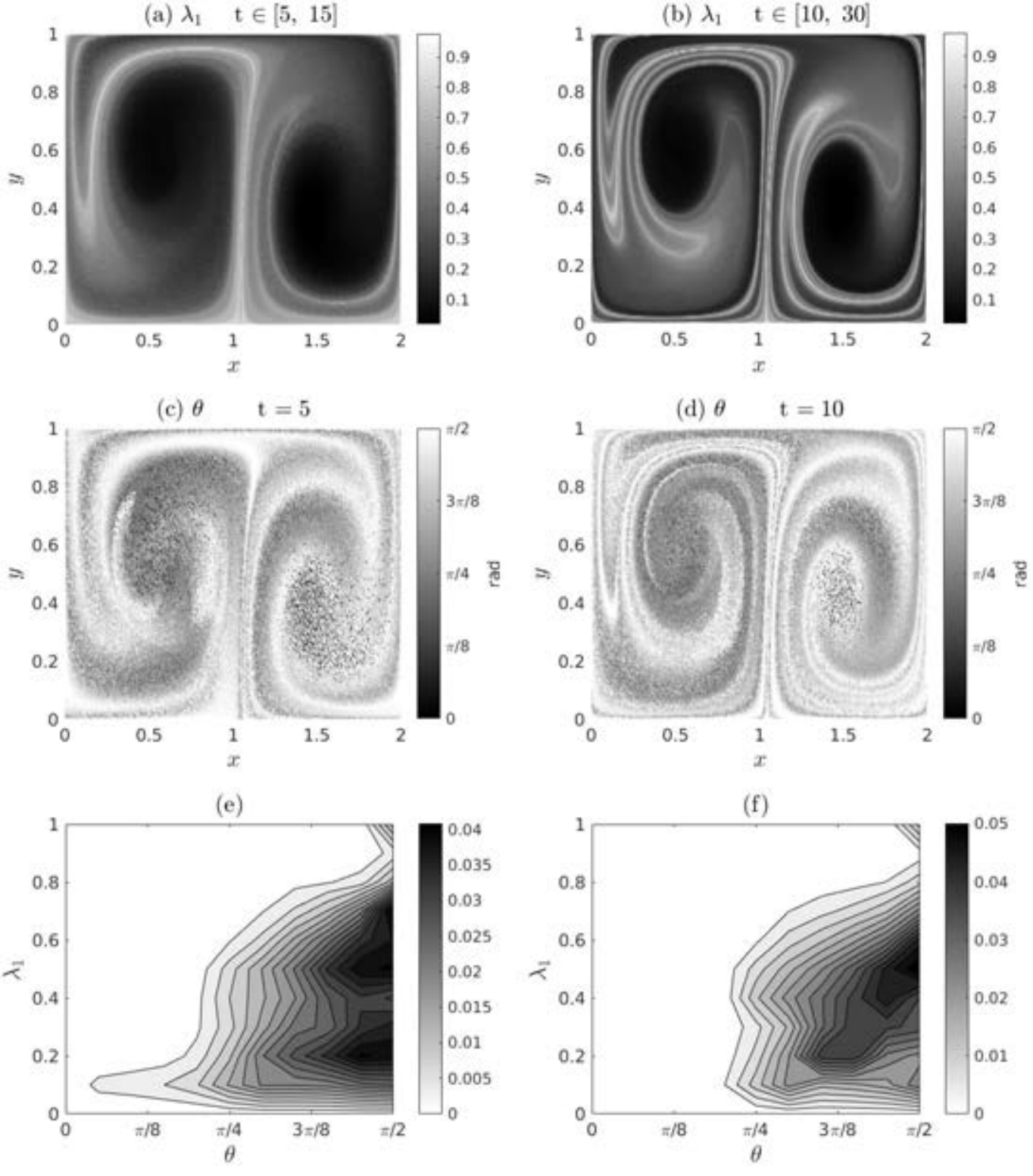}
\caption{ 
FTLEs,  CLVs and joint probability plot for the experiments DV1 (left panels) and DV2 (right panels) of the double gyre system. The first column is referred to the experiment DV1 and the second column to DV2. Panels (a) and (b) show the maximum FTLE fields computed from \eqref{avel}. Panels (c) and (d) show the angle between the CLVs for the two experiments. Panels (e) and (f) show the joint probability between FTLEs and angles shown in the previous panels. }
\label{fig:4}
\end{figure}

We consider two numerical experiments,  \emph{DV1} and  \emph{DV2},  in  which the CLVs are computed in two different time intervals with a non-null intersection. During \emph{DV1} we compute the CLVs in the interval $t=[5,\,\,10]$ and  during \emph{DV2} for $t=[10,\,\,20]$. This is important, as will be shown in this section, to highlight that $\theta$ is a quantity that depends on a particular time instant and not on the time interval considered for its computation, as long as the interval considered is sufficiently long.
Figure \ref{fig:4} shows a comparison between the angle $\theta$  and the FTLEs for this system. The first column is relative to  \emph{DV1}, while the second column is relative to \emph{DV2}.  
The $\lambda_1$ fields, the maximum FTLEs, are shown in Figures \ref{fig:4}a and \ref{fig:4}b  for \emph{DV1} and  \emph{DV2} respectively. The regions characterized by white color correspond to  maximum exponential growth rate. Ridges of these regions correspond to the LCSs of the system according to the definition by Shadden et al \cite{Shadden2005}. It is interesting to notice how, although in Figure \ref{fig:4}b the ridges of the FTLEs are more developed, the $\lambda_1$ field  converges to the same structures, and the maximum values do not change. The FTLEs give an overall information of the stretching and folding over the whole time interval considered, they can not say anything about the state of the coherent structures at a particular instant of time.
Figures \ref{fig:4}c and \ref{fig:4}d show the hyperbolicity field $\theta$  computed for a particular time instant.  The white colour here represents the maximum values of the hyperbolicity of the system, which means $\theta=\pi/2$, and highlight possible HCCSs as defined in \ref{defHCCS}. Although the  FTLEs and $\theta$ fields exhibit  similar structures it is interesting to point out that the ridges of the FTLEs do not necessarily correspond to the regions of maximum hyperbolicity. 
Near the central regions of the two vortices, the hyperbolicity field has a noisy appearance, due the fact already discussed that in there the expansion and contraction directions are not  well defined, and the  CLVs become tangent to each other. Notice that this effect does not influence the detection of hyperbolic regions.  
 Figures \ref{fig:4}e and \ref{fig:4}f  highlight the differences between FTLEs and the $\theta$ fields shown respectively for the two experiments  \emph{DV1} and  \emph{DV2}. As already mentioned, high values of the FTLEs do not necessarily correspond to high values of the $\theta$ field. 
 The joint probability plots underline the fact that in general there is not a one to one  relation between the $\theta$ and FTLE fields.  In particular, small values of FTLEs are related to the broadest range of possible values for the angle between CLVs.  This can be understood considering that the FTLEs measure the exponential growth rate of divergence of nearby trajectories. Where the FTLEs are smaller, the expansion and contraction directions are not well defined and a broader range for the possible values of the angle is obtained.  Although there is not a clear relation between the two quantities  it is interesting to point out that, for this example,  the maximum values of FTLEs correspond to a reduced range of possible values of $\theta$, and for the  \emph{DV2} experiment  there is a univocal correspondence. 
 The joint probabilities are asymmetric and their peaks towards the highest values of $\theta$ highlight  the fact that the system is hyperbolic. In this case, in fact,  it is possible to find, almost for every point of the domain, well defined expansion and contraction directions. It is interesting to note that the joint probability of Figure \ref{fig:4}f is less asymmetric with respect to the correspondent joint probability in Figures \ref{fig:4}e.
 
\begin{figure}
\centering
\includegraphics[clip,width=0.9\columnwidth]{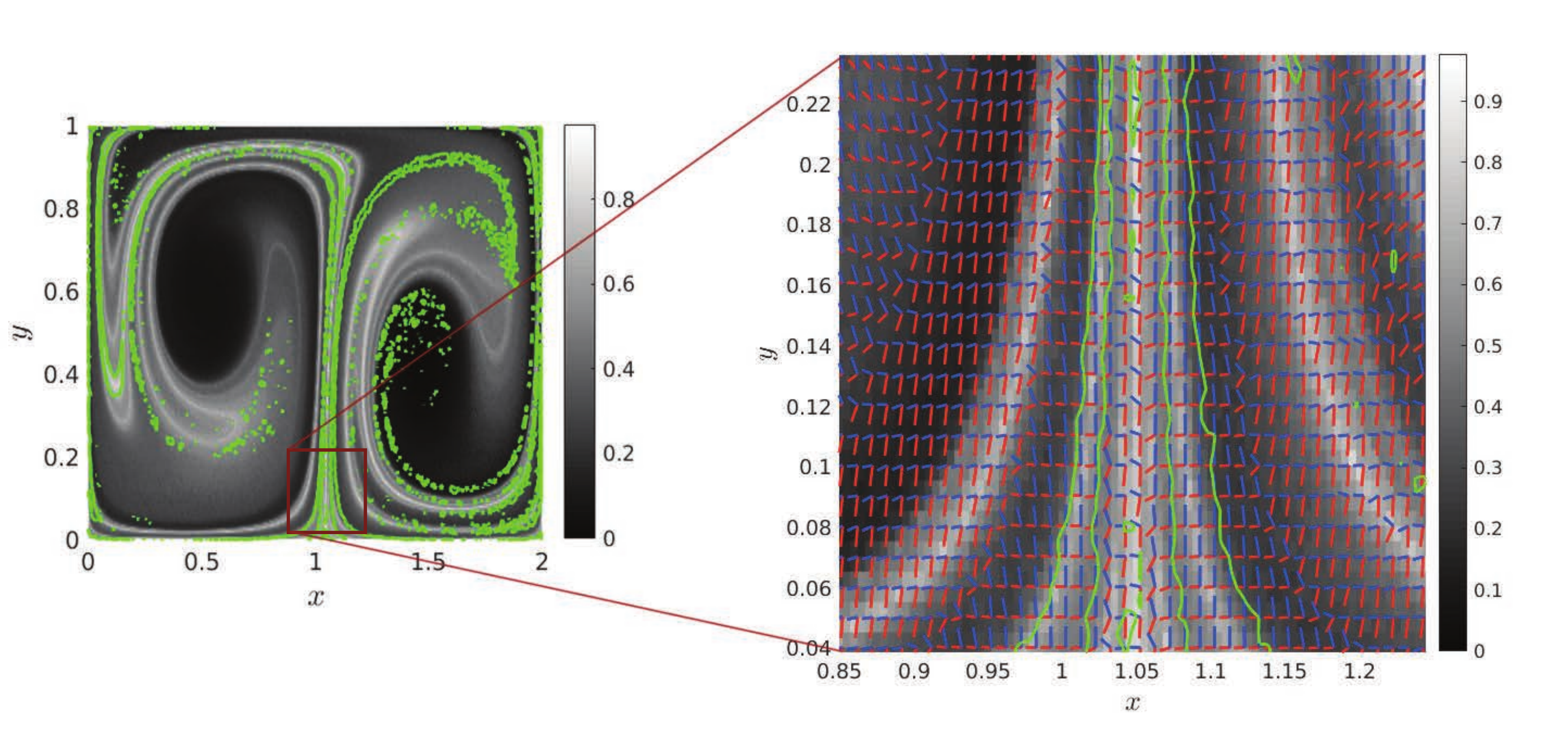}
\caption{Comparison between the HCCSs at time $t=5$, green lines, computed for the  experiment  \emph{DV2} and the FTLEs field shown in Figure \ref{fig:4}b. In the right panel there is a zoom of a portion of the domain in which are also shown the CLVs characterizing that region. Blue direction is contracting while red is expanding. Since the contour in this region is aligned with the second CLVs, the blue one, the character of the HCCSs here is repulsive.}
\label{fig:5} 
\end{figure}

   
Figure \ref{fig:5} compares the HCCSs at time $t=10$ for the experiment  \emph{DV2}, found with the definition \ref{defHCCS}, and the FTLEs field shown in Figure \ref{fig:4}b. In the right panel there is a blow up of a region containing HCCSs. These pictures show that the instant structures highlighted by the HCCSs do not always correspond to the ridges of the FTLEs. The contours of the regions characterized by $\theta=\pi/2$ define the shapes of  HCCSs, but  to understand their attractive or repelling nature it is necessary looking at the CLVs that characterized those regions. The right panel of Figure \ref{fig:5} shows the repelling nature of the HCCSs in the zoomed region, since the contour is aligned with the second CLVs, ${\bf w}_2$, that define the contraction direction. Every particle of the tracer near these HCCSs will tend asymptotically to get away from them.

%
\begin{figure}
\centering
  \includegraphics[scale=1]{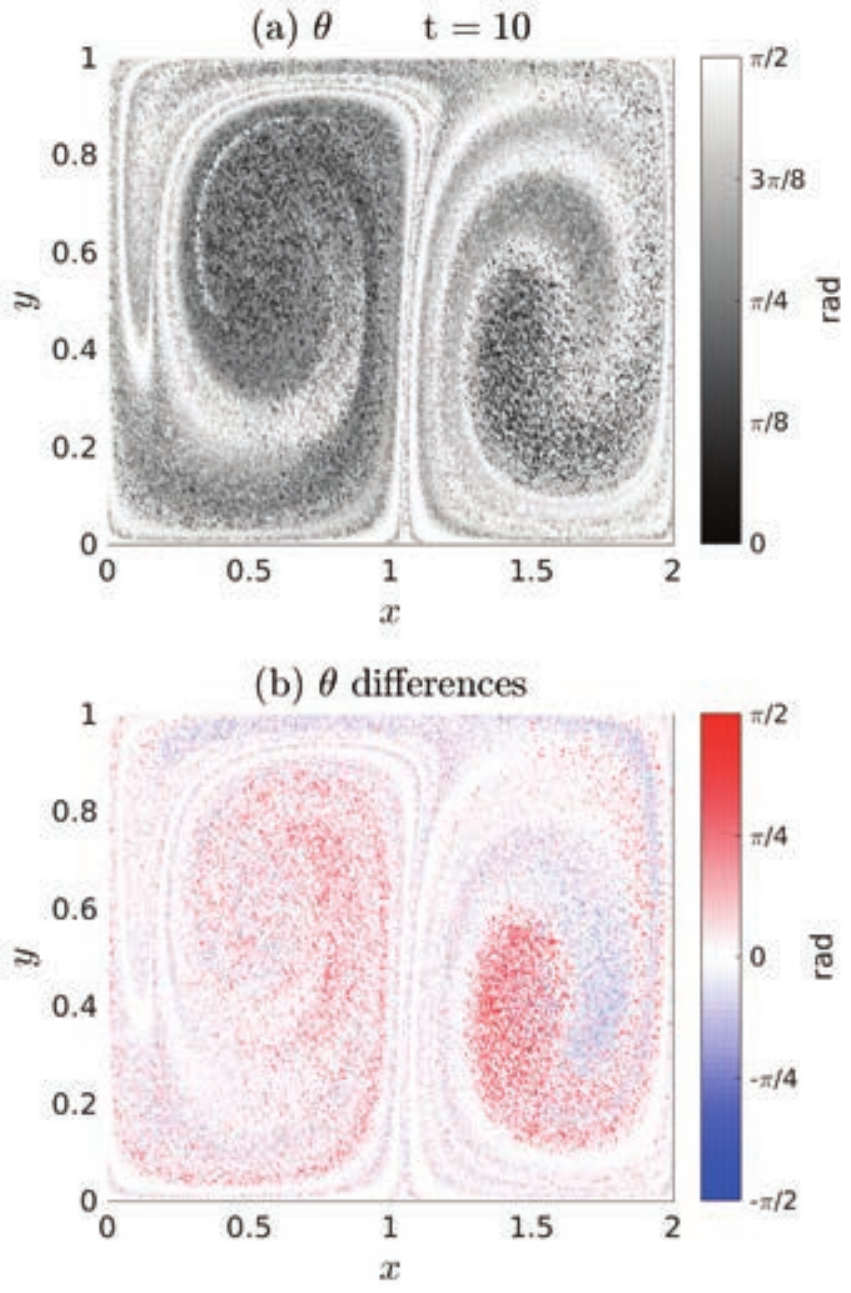}
\caption{Comparison of the angle between CLVs, $\theta$, computed for the two experiments DV1 and DV2 at the same time $t=10$. Panel (a) show the evolution of the angle obtained in \emph{DV1} at $t=10$. Panel (b) shows the difference between the angle computed at the beginning of  the experiment \emph{DV2} and the one computed at the end of the interval of the CLVs computation for the experiment \emph{DV1}.  }
\label{fig:6}
\end{figure}
Figure \ref{fig:6} compares the evolution of $\theta$ field in \emph{DV1}, computed at $t=10$, with the angles computed at the same time for  the experiment \emph{DV2}.
The region of maximum hyperbolicity
that appears in Figure \ref{fig:6}a, corresponds to the same region that is found in Figure \ref{fig:4}d. The difference between the two fields (Figure \ref{fig:6}b) shows large values in regions with low hyperbolicity, and values close to zero for $\theta = \pi /2$ implying a convergence of the $\theta$ field. It should also be noted that the structure  highlighted by the maximum hyperbolicity in the left part of the domain is pretty similar to the shown in Figure \ref{fig:4}a. Once again, this underline that the FTLEs field gives an overall information and not an instantaneous one. The HLCSs found by using the FTLEs are, by definition, strongly related to the time interval chosen for the study, while the HCCSs, which correspond to  instantaneous structures of maximum  hyperbolicity,  are independent of the time interval considered for their computation, as shown in these two experiments. The short-term behavior of passive tracer analyzed by HLCSs is strongly influenced by the time window used with respect to long-term behavior highlighted by HCCSs. It should also be noted that the HLCSs, once they have been found, are advected with the flow so as to ensure that they act as barriers for the passive tracer. However, this precludes the study of any new barriers that may arise in the flow in a time subinterval. These constraints are not present in the long-term study performed with HCCSs.

\begin{figure}
\centering
\includegraphics[scale=0.7]{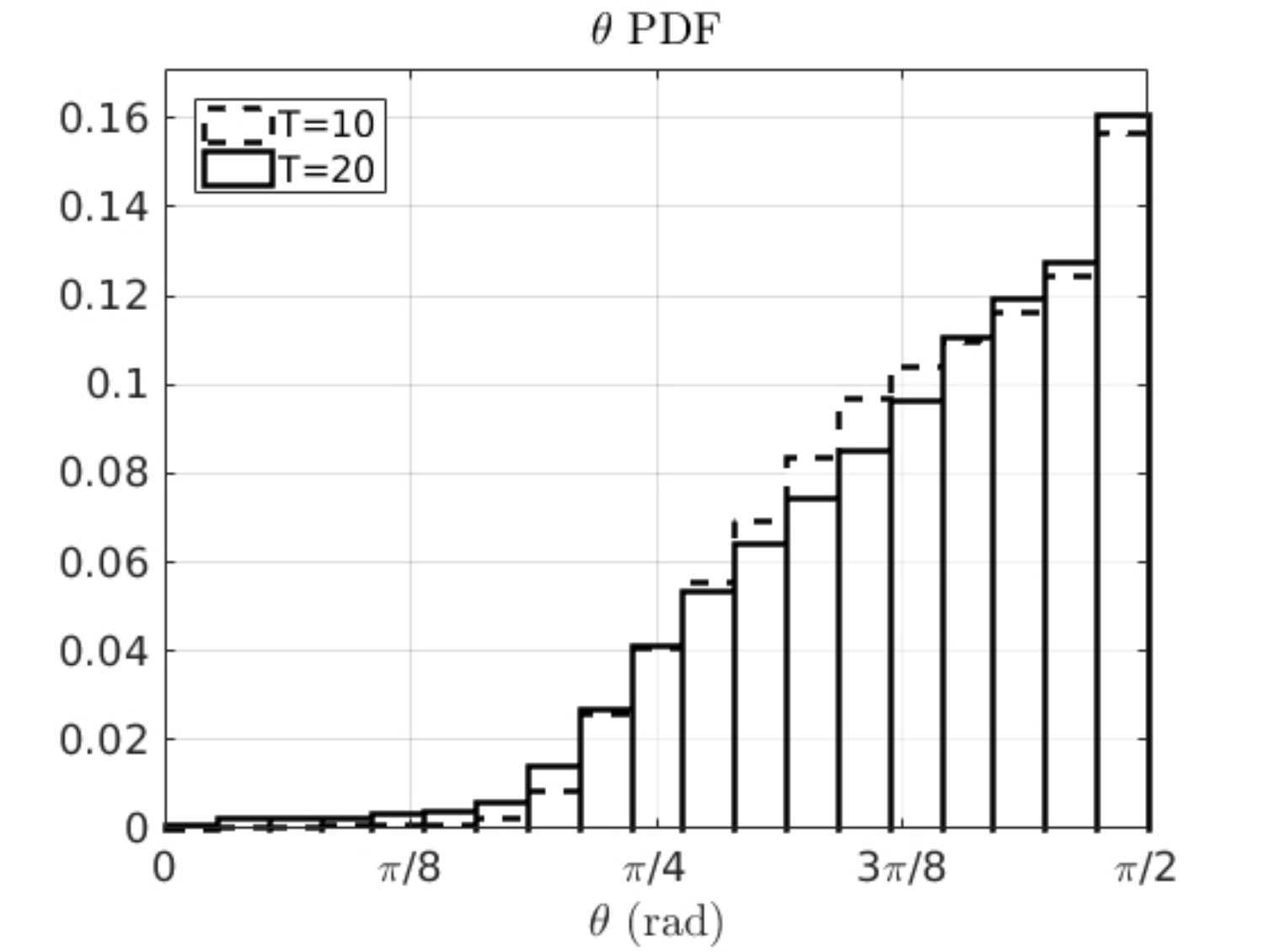}
\caption{PDF of $\theta$ for the \emph{DV2} experiment computed at $t=10$ (dashed line) and at $t=20$ (full line).}
\label{fig:7}
\end{figure}
\begin{figure}
\centering
\includegraphics[scale=0.6]{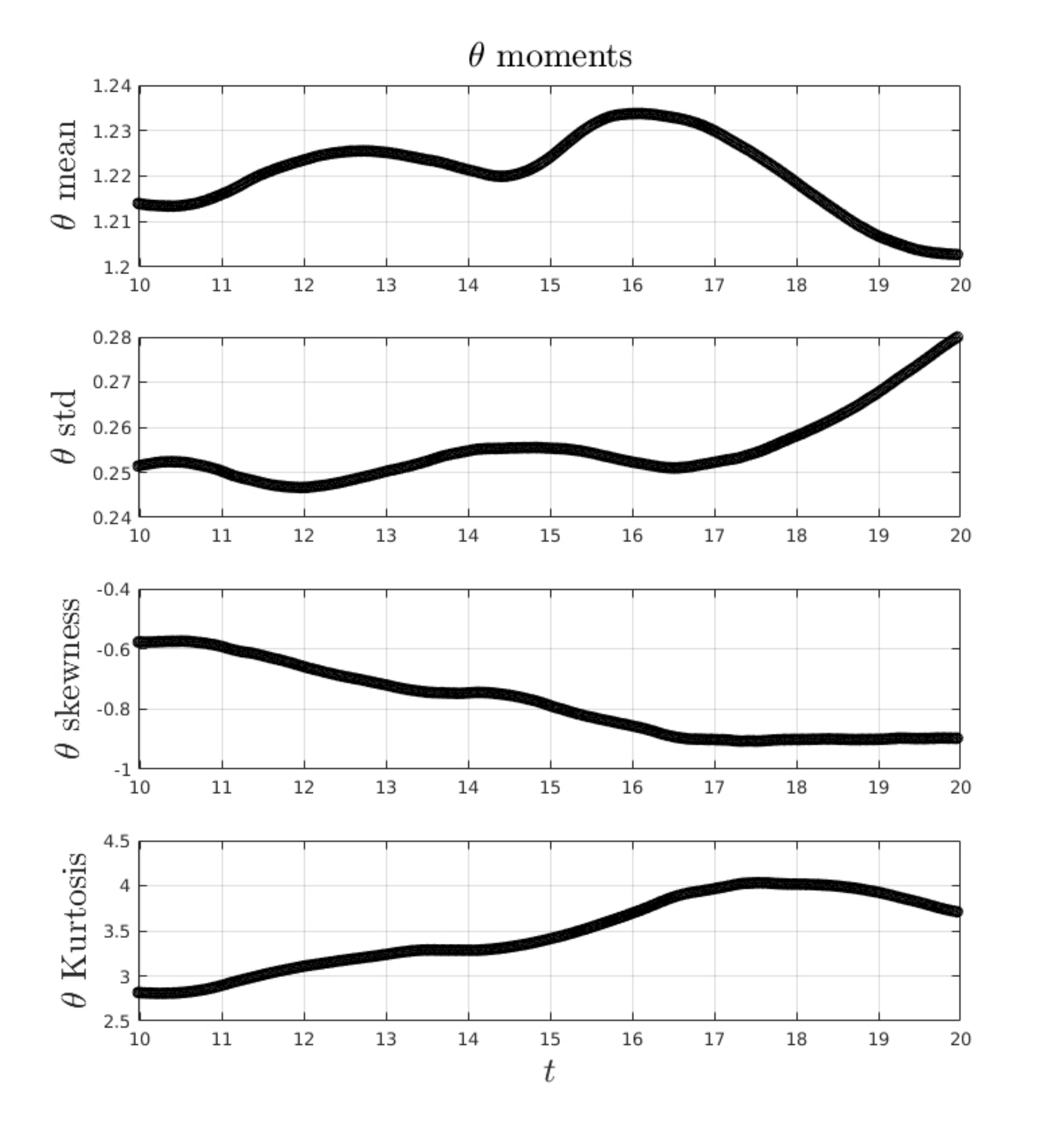}
\caption{Evolution of the first four moments of the PDF for $\theta$ for the DV2 experiments in the interval $T2=[10\,\,20]$. }
\label{fig:8}
\end{figure}
It is also interesting to consider the evolution of the PDF  of the hyperbolicity field, $P(\theta,t)$. If the distribution  is peaked around $\pi/2$,  the expansion and contraction directions are almost everywhere perpendicular between each other. 
 If the distribution is peaked around zero,  the expansion and contraction directions are almost tangent everywhere. If the distribution is flat there is not a clear correlation between the expansion and contraction directions.
 Figures \ref{fig:7} shows the PDF of $\theta$  computed for the initial and final time of CLVs computation for the experiment \emph{DV2}.
 The final PDF (full line) highlight an increasing of points with  hight hyperbolicity,  and an increment of points in the tail of the distribution near zero, with respect to the PDF computed at the beginning of the interval (dashed line). During the evolution in the interval $t=[10,\,\,20]$ the distribution becomes more asymmetric.
 The change in shape is visible also looking at  the first four moments of the PDF computed at every time step in this time interval (Figure \ref{fig:8}).
 All the moments of the distribution display only small changes in time during the evolution.  This time  interval is characterized by a   mean value of $\theta$ oscillating in the range $[1.20,\,\,1.23]$. A monotonic change in the mean, possibly given by mixing, is visible after  $t \approx 16$. It should be noted however that the range of this change is very small. The s.t.d. of the angles distributions is characterized by values that change within the interval $[0.25,\,\,0.28]$. After $t \approx16$ the s.t.d. shows a monotonic increase, as a signature that, if the change in time is given by mixing, this must be of adiabatic nature and thus not resulting in a homogeneization of the field. The third moment of the PDF decreases in time from $-0.57$ to $-0.91$, showing thus a negatively skewed distribution. Finally, the kurtosis initially increases from $2.81$ to $4.03$, and then shows a slight decrease in values, generally indicating a less flat distribution than the normal distribution. 
 Although the motion of the velocity field is periodic, the motion of the tracers in this flow is chaotic. This is the reason why we do not see in this interval a periodicity in the moments of the PDF for the variable $\theta$.

\subsection{\label{sec3:2-3}Bickley jet}

The Bickley jet is an idealized model of a jet perturbed by a Rossby wave \cite{Negrete1993,Vera2010,Onu201526}.  The velocity field is given in terms of the stream function, $\psi(x,y,t)$, that can be decomposed as the sum of a mean flow $\psi_0(x,y,t)$ and a perturbation $\psi_1(x,y,t)$
\begin{equation}
\psi(x,y,t)=\psi_0(x,y) + \psi_1(x,y,t),
\end{equation}
where
\begin{equation}
\psi_0(x,y) = c_3 y -UL_y\tanh\left(\frac{y}{L_y}\right) +\\ \epsilon_3 U L_y \text{sech}^2\left(\frac{y}{L_y}\right)\cos(k_3 x),
\end{equation}
and
\begin{equation}
\psi_1(x,y,t) = UL_y \text{sech}^2\left(\frac{y}{L_y}\right)\mathcal{R}e\left[ \sum_{n=1}^{2}\epsilon_n f_n(t)e^{ik_nx}\right].
\end{equation}
Following the work by Onu et al. \cite{Onu201526}, the forcing is chosen as a solution that runs on the chaotic attractor of the Duffing Oscillator
\begin{subequations}
\begin{eqnarray}
&&\frac{d\varphi_1}{dt} =\varphi_2\\
&&\frac{d\varphi_2}{dt} =-0.1\varphi_2-\varphi_1^3+11\cos(t),\\
&&f_{1}(t)=f_{2}(t)=2.625\, 10^{-2}\varphi_1(t/6.238\times10^5).
\end{eqnarray}
\end{subequations}

From now on the time will be scaled with the quantity $L_x/U$.
The streamline of the flow at $t=1.89$ are shown in Figure \ref{fig:2}c.

As for the double gyre, we show two experiments, $BJ1$ and $BJ2$, for which the end of the CLVs computation in the first experiment correspond with the first CLVs computation for the second experiment.
\begin{figure}
\centering
\includegraphics[scale=1]{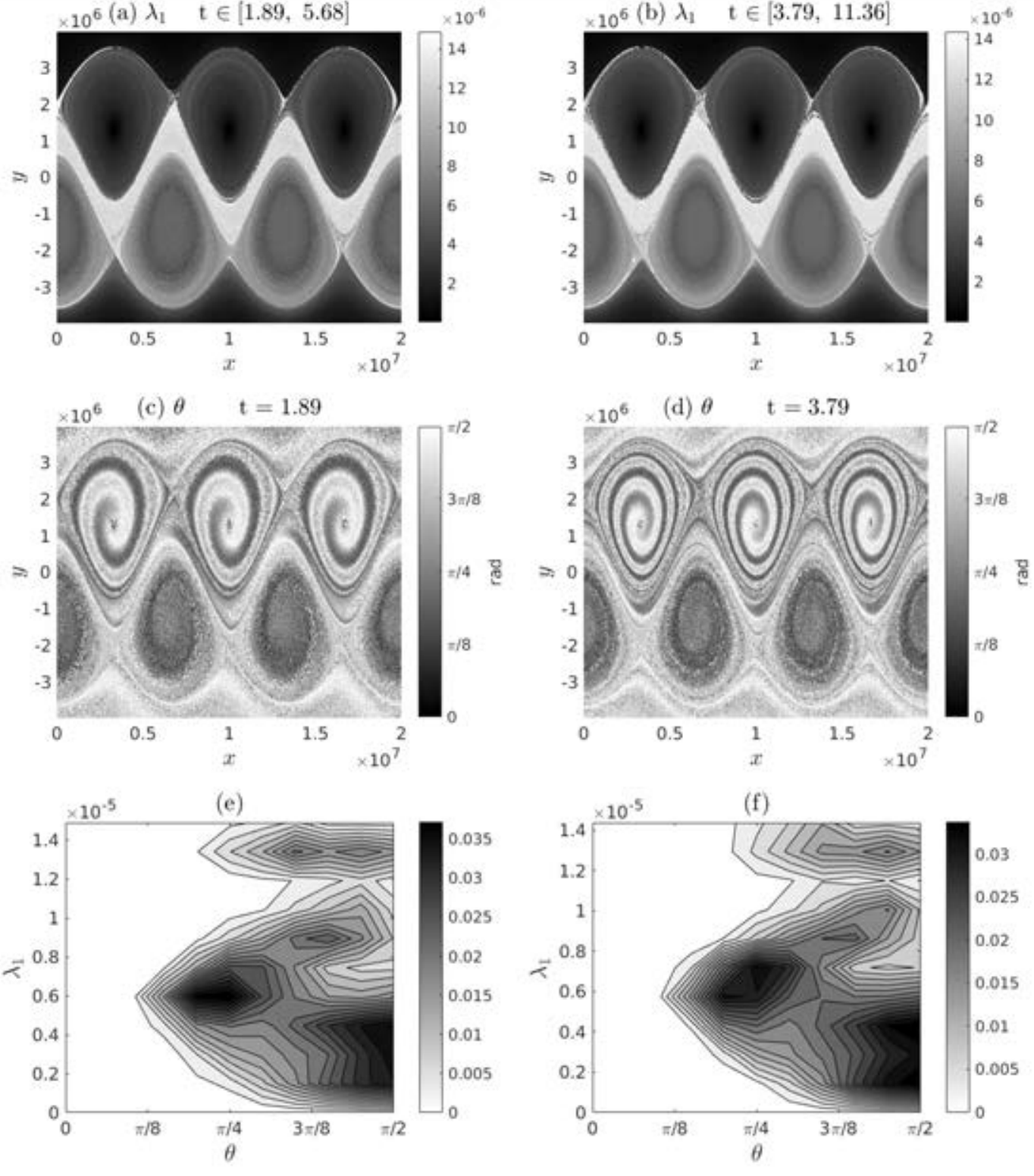}
\caption{As in Figure \ref{fig:4}, but for the two experiments \emph{BJ1} and \emph{BJ2}.}
\label{fig:9} 
\end{figure}
The first column of Figure \ref{fig:9} shows the analysis for \emph{BJ1} and the second column for \emph{BJ2}. Figures \ref{fig:9}a and \ref{fig:9}b, show the field of the maximum FTLEs, $\lambda_1$. Results show the convergence for the FTLE fields. 
Maximum values of $\lambda_1$ are reached in the  central jet and around the boundaries of the vortices.  Figures \ref{fig:9}c and \ref{fig:9}d show the angle between CLVs, where the white color indicates $\theta=\pi/2$. The $\theta$ fields clearly provide more details of the flow patterns than the FTLE fields. In agreement with the FTLE fields, values of $\theta$ close to $\pi/2$ are present in narrow bands in the central jet, in the center and  around the vortices. The angle field shows however that not all the central band of the FTLEs field represents an HCCSs.  The $\theta$ fields show also spiraling patterns within the vortices, which are instead not visible in the FTLE fields. The spiral patterns are particularly evident in the upper vortices, where the perturbation $\psi_1$ acts as a positive feedback to the flow. In the bottom vortices, the perturbation acts instead to weaken the flow. Figures \ref{fig:9}e and \ref{fig:9}f show the joint probability plots of the angles and the FTLEs just presented. These plots show that there is no clear relation between FTLEs and angle.
For a given FTLE correspond many values of the angle between CLVs. As for the double gyre example, smaller values of the FTLEs correspond to the maximum range of possible values for $\theta$. However, differently from the double gyre, also high values of FTLEs  correspond to a broad range of $\theta$ values. This can be explained considering that almost all the points corresponding to the highest values of FTLEs are contained in the central jet. In this region, the FTLEs  converge to the same value, independently from the space position along the jet or  time. Due to this convergence we do not have much information about smaller structures that can characterize the system in the jet region as instead shown by the $\theta$ field. For this reason, the joint probability plots do not exhibit  a reduced range of $\theta$ values in correspondence of the highest FTLEs. The comparison between the \emph{BJ1} and \emph{BJ2} integrations shows that for the Bickley jet  both the PDFs of the  FTLEs and $\theta$ field are more stationary in time with respect the double gyre.
\begin{figure}
\centering
\includegraphics[scale=0.4]{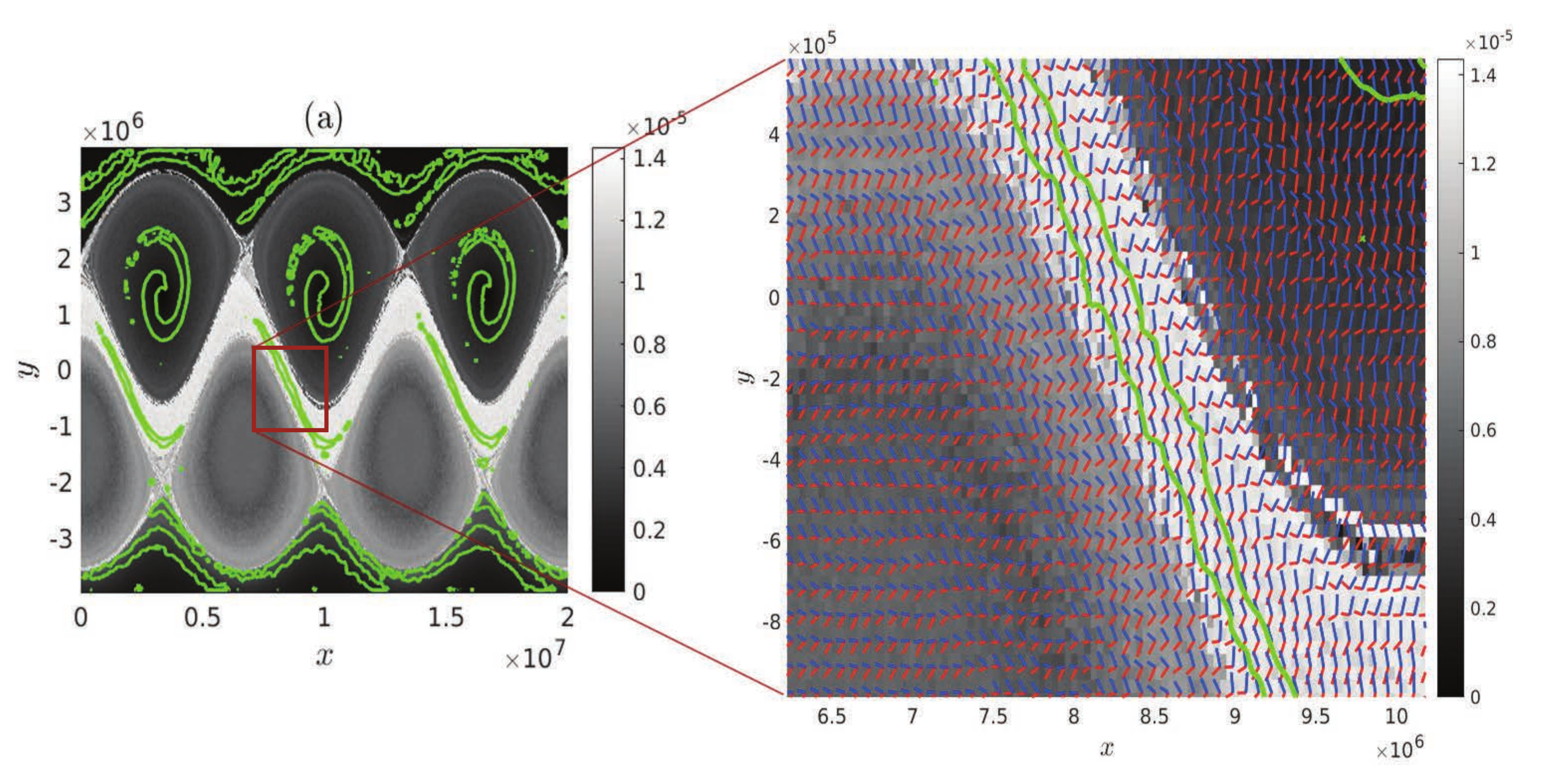}
\caption{ As in Figure \ref{fig:5}, but for the two experiments \emph{BJ1} and \emph{BJ2}.}
\label{fig:10}
\end{figure}
The left panel of Figure \ref{fig:10} shows the superposition of HCCSs computed for $t=3.79$ for the experiment \emph{BJ2}, green lines, and the FTLEs computed in the same experiments (Figure \ref{fig:9}b). This figure remarks the fact that the central jet is not completely a HCCSs. HCCSs are also present at the border of the vortices and in the center part of the upper vortices. The right panel of Figure \ref{fig:10} show an enlargment into a portion of the central jet containing HCCSs and the CLVs in that part of the domain. In this picture it is possible to appreciate the repelling nature of the HCCSs looking the alignment of the contour with the second CLVs. It is interesting to point out that, looking at the FTLEs field, it is not possible to see ridges in the central jet. In this case it is not possible to find LCSs, if we use definition of LCSs based on the FTLEs field \cite{Shadden2005, Haller2011}, and this once again remark the difference between the HCCSs and LCSs.

\begin{figure}
\centering
\includegraphics[scale=1]{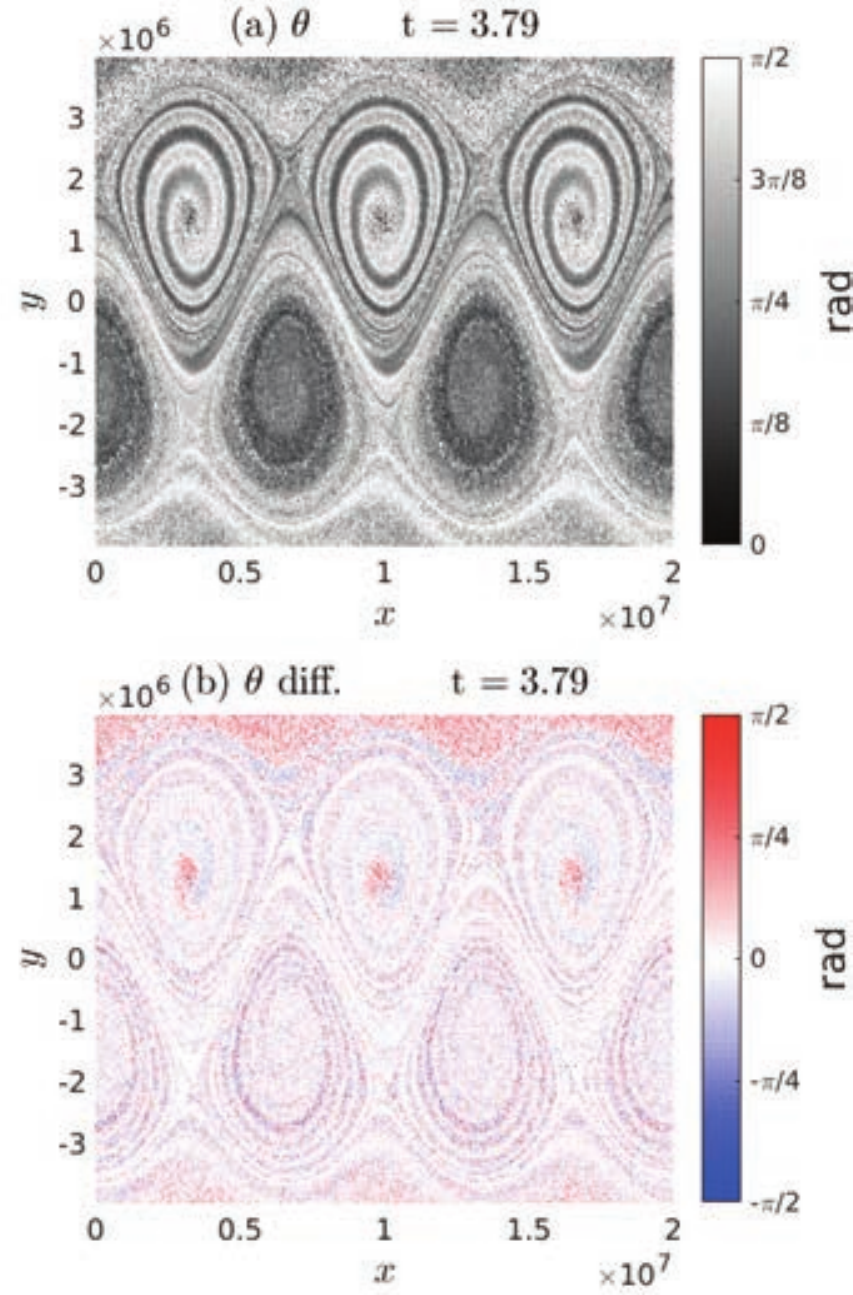}
\caption{As in Figure \ref{fig:7}, but for the \emph{BJ2} experiment.}
\label{fig:11} 
\end{figure}

Figure \ref{fig:11} shows the comparison between the angles  at the end of the CLVs computation interval of the experiment $BJ1$ with the one computed  at the beginning of the CLVs calculation interval for the $BJ2$ experiments. As for the double gyre, the comparison between Figure \ref{fig:11}a and Figure \ref{fig:9}d shows that regions characterized by $\theta = \pi /2 $, i.e. the HCCSs, have the same structure. The difference between the two fields (Figure \ref{fig:11}b), shows that these regions are characterized by smaller errors. In contrast, the regions with low hyperbolicity of Figure \ref{fig:11}a appear to be more noisy with respect to the same regions computed for the $BJ2$ simulation (Figure \ref{fig:9}d). In agreement with this, the difference between the two fields shows that the low hyperbolicity regions are characterized by larger errors (Figure \ref{fig:11}b).
%
\begin{figure}
\centering
\includegraphics[scale=0.6]{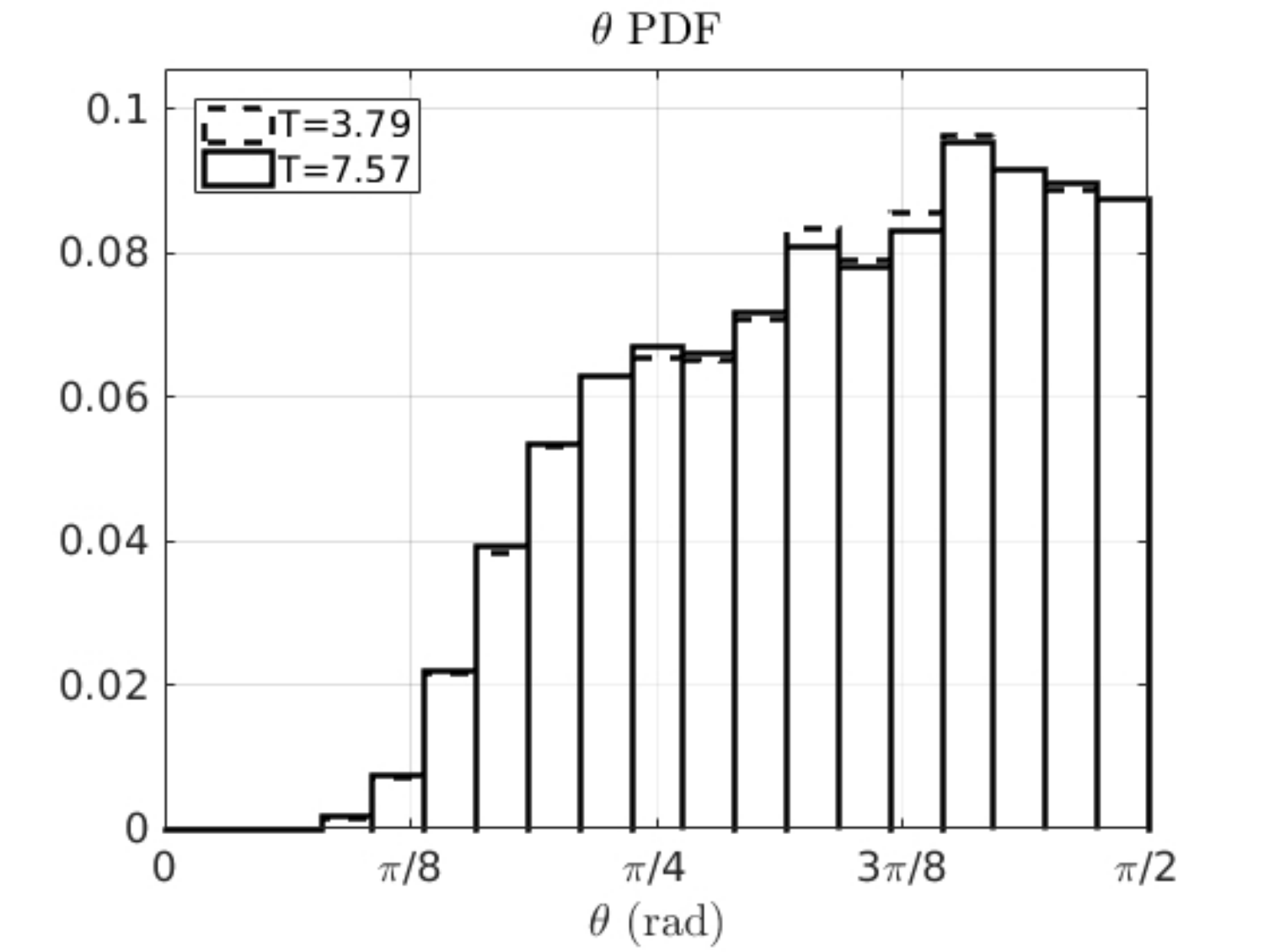}
\caption{As in Figure \ref{fig:7}, but for the \emph{BJ2} experiment.}
\label{fig:12} 
\end{figure}
Figure \ref{fig:12} shows the PDF of $\theta$ for the experiment \emph{BJ2} at the beginning and at the end of the CLVs computational interval. Results show that the PDF of the angle for this this interval remains stable. This is confirmed by the analysis of the moments of the distribution (not shown), which appear to be constant in time.  With respect to the double gyre, the PDF for the Bickley jet is more flat indicating the presence of a larger number of points in which the CLVs are not orthogonal.

\section{\label{sec4:1}Conclusions}
We have here proposed a new definition and a new computational framework to determine hyperbolic structures in a two dimensional flow based on Covariant Lyapunov Vectors. CLVs are covariant with the dynamics, invariant for temporal inversion and  norm independent. These vectors are the natural mathematical entity to probe the asymptotic behaviour of the tangent space of a dynamical system.
 All these properties allow an exploration of the spatial structures of the flow, which can not be done using the Lyapunov Vectors bases due to their  orthogonality.

CLVs are related to the contraction and expansion directions passing through a point of the tangent space, and the angle between them can be thus considered as a measure  of the hyperbolicity of the system. This information can be summarized in a scalar field, the angle $\theta$, between the CLVs referred to the initial grid conditions, and used to define hyperbolic structures. The structures identified with the isolines of this field, characterized by $\theta=\pi/2$, are called   Hyperbolic Covariant Coherent Structures. These patterns  are the most repelling or attracting pointwise, \textit{in terms of the  asymptotic behaviour of tracer particles}, with respect to nearby structures at a given time. In terms of practical applications, this has important consequences, as it will provide an indicator for the long time transport of passive tracers such as for example oil spills in the ocean.

CLVs, and the correspondent $\theta$ field, have been computed for three numerical examples to compare  how the behaviour of the particles tracer near HLCSs, highlighted by the FTLEs, can change asymptotically in time. The three examples include an Hamiltonian autonomous system, and two non-autonomous systems that are bounded or periodic. For all these examples it is possible to compute CLVs, HCCSs, and compare them with the HLCSs.
Since the FTLEs tend to converge to LEs and lose their dependence on the initial conditions the angle between the CLVs could give more detailed information about possible structures that can emerge from the flow.  This feature  has been highlighted in particular for the Hamiltonian autonomous system, in which $\theta$ is able to detect the central barrier in contrast with the FTLEs field, and in the Bickley jet in which the FTLEs converge to the same value in the jet region and it is not possible to see any kind of particular finest  structure.
The use of $\theta$ provides information on the structures appearing at each time of the evolution of the flow, and the three examples underline that not always the HCCSs correspond to HLCSs and vice versa. So, particles tracer, such as chlorophyll or oil in water, can be maximally attracted or repelled by some HLCSs, but if we consider a different time interval and in particular the asymptotic behaviour of these particles, we can obtain a  distribution that is completely different. Note that, in practice, the asymptotic time length can be considered as the time taken by two random initial basis to converge to the same BLVs basis. 

It should be noticed that, while  no fluxes can be present across the HLCSs
the same does not necessary hold for all the structures appearing characterized by $\theta = \pi / 2$. HCCSs can be found for every instant of time but they give the asymptotic information about the behaviour of the particles  tracer near the HCCS at that instant of time and, for this reason, the zero flux requirement of the HLCSs is not necessary. \textit{HCCSs are not necessarily barrier, their meaning is different from the one of HLCSs}.

For the three examples considered we have also computed HLCSs with the geodesic theory using the LCS tool \cite{Onu201526} (not shown). The results  are in agreement with the discussion above. Looking at asymptotic time it is still possible to find particular structures emerging from flow, but clearly these structures not necessarily correspond to HLCSs computed for a particular time interval.

For the two non-autonomous systems we have considered the evolution of the PDFs of the $\theta$ field and the evolution of its first four moments.
For the Bickley jet the probability distribution of the angle is stationary in time, and so its moments, but for the double gyre it is possible to appreciate a small variation in time for the PDF. The information deriving by the evolution of the  $\theta$ field, related to the variation of the strength of the hyperbolicity field, could be used to characterize the dynamical mixing of the system.

 Finally, future studies will have to address the detection of hyperbolic structures beyond analytical systems, i.e. for two-dimensional turbulent flows. This will be particularly interesting for flows at the transition between balance and lack of balance \cite{badin2011lateral,badin2014role,badin2013ssg,ragone2016study}, where the detection of HCCSs can shed light both on the structure of mixing and on the forward cascade of energy to dissipation, or the lack of thereof.  In particular, the evolution of the moments of the PDF of $\theta$ could be used to define an index of dynamical mixing for the system under study. Particularly interesting will be the behaviour of the HCCSs in presence of intermittency. We can conjecture that in particular cases, such as e.g. the merging of two vortices, the instantaneous structures underlined by the HCCS can give reliable information of the asymptotic tracer dynamics, that is the dynamics after the merging event.  This is however left for future studies.

\vspace{6pt} 



\acknowledgments{We would like to thank two anonymous referees for comments that helped to improve the manuscript. This study was partially funded by the research grant DFG1740. GB was partially funded also by the research grants DFG TRR181 and  DFG BA 5068/8-1. The authors would also like to thanks Sergiy Vasylkevych and 
Sebastian Schubert  for interesting discussions on the subject.}

\appendix

\section{Description of the algorithm}
\label{sec3:1-0}

The algorithm for the identification of CLVs is based on five different phases:

\begin{enumerate}
\item \emph{Initialization} (\emph{T1}): this preliminary step is used to find the initial  backward Lyapunov vector bases  $\{ {\bf l}^-_1(t),\,\, {\bf l}^-_2(t)\}$ for a whole set of initial conditions \cite{Benettin1980}.  A set of initial condition $\{{\bf x_0}\}\in D$ and  two sets of initial orthonormal random bases are defined in  the tangent spaces at every point  at time $t_0$. The second set is necessary to check the convergence to the backward Lyapunov vectors bases. 
The initial conditions and the random bases are evolved respectively with \eqref{ds} and \eqref{pd} until the convergence of the two bases is reached with the desired accuracy. The convergence toward the Lyapunov vectors is typically exponential in time \cite{Ginelli2007,Goldhirsch1987}.
 At every time step, for every initial condition, the evolved vectors are stored as column of a matrix that is decomposed with a QR decomposition. The last passage is implemented in order to find the new orthogonal basis at every time step, and the upper triangular matrix containing the coefficients that allow to express the old basis in terms of the new one.

\item \emph{Forward Transition} (\emph{T2}):   the backward  Lyapunov bases are evolved from time $t$ to $t'$. The evolution is done with the help of \eqref{ds}, \eqref{pd} and the $QR$ decomposition for every evolution step. We indicate with ${\bf X}(t_k)$ the matrix which columns contain the new bases at time $t_k$, and with  ${\bf R}(t_{k-1},\,t_k)$ the correspondent upper triangular matrix.
 During this step both the local Lyapunov bases and upper triangular matrices are stored.
The diagonal elements $({\bf R}(t_{k-1},\,t_k))_{ii}$ of the upper triangular matrices give information about the local growth rates of the bases vectors at a given time $t_k$, and they are used to compute the FTLEs as a time average
\begin{equation}
\lambda_i = \frac{1}{t'-t}\sum_{k=0}^{N-1}\log ({\bf R}(t_{k-1},\,t_k))_{ii},
\label{avel}
\end{equation}
where $N$ time step are considered between $t$ and $t'$.
If the  LEs exist, (\ref{avel}) will converge to them for a sufficiently long evolution. It is worth to note that sometimes the FTLEs are computed using a different method, that is using the so called Cauchy Green Tensor (CGT) defined as
\begin{equation}
{\bf G}(t, t')={\bf F}(t, t')^{\top}{\bf F}(t, t').
\label{CGT}
\end{equation} 
This operator is also known as deformation tensor, whose eigenvalues $\mu_i(t_0, t)$, and eigenvectors $\boldsymbol\xi_i(t_0, t)$, satisfy
\begin{subequations}
\begin{align}
&{\bf G}(t, t')\boldsymbol\xi_i(t, t')=\mu_i(t, t')\boldsymbol\xi_i(t, t'),\\ 
&\mu_1(t, t')>\mu_2(t, t')>0, \\
& \boldsymbol\xi_1(t, t')\,\bot \, \boldsymbol\xi_2(t, t').
\end{align}
\end{subequations}
From a geometric point of view, a set of initial conditions corresponding to the unit sphere is mapped by the dynamics into an ellipsoid, with principal axis aligned in the direction of the eigenvectors of the CGT and with length determined by the correspondent eigenvalues. 
The eigenvalues of the CGT determine the FTLEs as
\begin{equation}
\lambda_i(t, t')=\frac{1}{2(t'-t)}\log (\mu_i(t, t')),
\label{lecgt}
\end{equation}
where the dependence on the starting position has here been suppressed.
This second method for the computation of the FTLEs exhibits some problems, for example, if just one  finite local  growth rate is taken into account considering a large time interval, (\ref{lecgt}) teds to zero and not to the LEs. The first method should be preferred.

\item \emph{Forward Dynamics} (\emph{T3}):  in this step the trajectories and the bases are further evolved from time $t'$ to time $t''$ using \eqref{ds} and \eqref{pd}. This time interval should grant the convergence, during the backward dynamic, to the CLVs. During this step only the upper triangular matrices  are stored and are used to continue the computation of the FTLEs using (\ref{lecgt}).

\item \emph{Backward Transition} (\emph{T4}): in this step random upper triangular matrices are generated for every point of the grid, ${\bf C}$. These matrices contain the expansion coefficients of a set of two generic vectors (expressed as column of a matrix) in terms of the ${L^-}$ bases. Using the stored matrices ${\bf R}$ of  the step 3,  these matrices are evolved backward in time, until time $t'$, using the following relation:
\begin{equation}
{\bf C}(t_n)={\bf R}^{-1}(t_n,t_{n+1}){\bf C}(t_{n+1}){\bf D}(t_n,t_{n+1}),
\label{be}
\end{equation}
where $t_n$ and  $t_{n+1}$ are time step between $t'$ and $t''$. This method uses all the information contained in ${\bf R}$ and not just the diagonal part of the matrix. The ${\bf D}$ diagonal  matrices contain the column norm of 
${\bf C}$. Using Eq. \eqref{be} it is possible to show \cite{Ginelli2007} that the generic vectors chosen will be aligned with the CLVs.
  Note that when a trajectory passes close to tangency of an invariant  manifold, the matrices $C$ can be ill-defined, and so a little amount of noise on the diagonal element of these matrices, or an average on the diagonal elements of the neighbor matrices is used to correct the problem.

\item \emph{Backward Dynamics} (\emph{T5}): in this final part of the algorithm,  \eqref{be} is used with the ${\bf R}$ matrices of the step 2 to evolve backward  the upper triangular matrices ${\bf C}$, from time $t'$ to time $t$. In this phase the  backward Lyapunov bases stored can be used to write the CLVs.  The matrix  containing in each columns the different CLVs at a given point in space and time, ${\bf W}$, can thus be written as
\begin{equation}
{\bf W}(t_n)={\bf C}(t_n){\bf X}(t_n).
\end{equation}

For a two dimensional system, the algorithm could be optimized making use of \eqref{relV}. One can follow the first step (\emph{T1}) of the previous algorithm to find the convergence toward the backward Lyapunov bases in the time interval $[t_0,\,\,t]$. After this first step it is possible to carry on the evolution of the vectors, as in the second step (\emph{T2}) during the time interval $[t,\,\,t']$, without saving the triangular matrices. In the same way it is possible to repeat the step (\emph{T1}) but for a backward evolution during the time interval $[t'',\,\,t']$ to find the forward Lyapunov bases and  then continue the backward evolution as in the step (\emph{T2}) during the time interval $[t',\,\,t]$.  At this stage it is possible to consider directly the first backward Lyapunov vectors and the second forward Lyapunov vectors in the time interval $[t,\,\,t']$ as the CLVs.  In this  algorithm there are just four steps and not five as in the one presented above, but for two times one has to consider the convergence step (\emph{T1}) that  is more time consuming in respect to the steps (\emph{T4}) or (\emph{T5}).  It should be noted that in this algorithm, the forward and the backward evolutions  could be done in parallel. The comparison between this algorithm and the one used for this study is left for future studies.

\end{enumerate}

\section{Technical details of the numerical examples}
\label{sec3:1-1}
\subsection{Hamiltonian system}
\begin{figure}
\centering
  \includegraphics[scale=0.6]{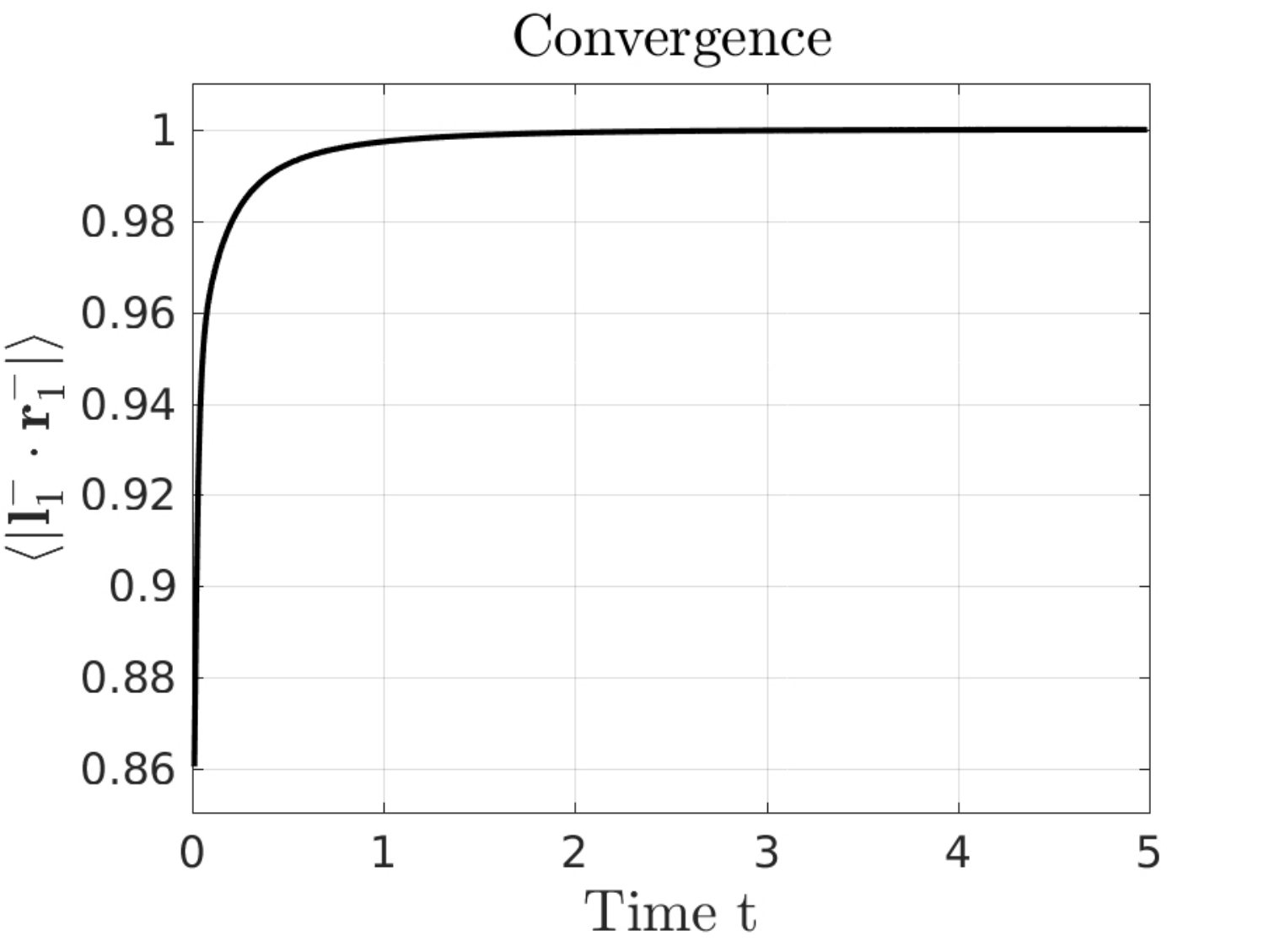}
\caption{Convergence of the averaged scalar product for the two random bases chosen for the initialization of the numerical algorithm for the Hamiltonian system \eqref{Heq}. }
\label{fig:13}
\end{figure}

For the simple Hamiltonian flow we  consider the domain $x=[0.2, \,\, 4]$, $y=[-0.5, \,\, 0.5]$,  a resolution of $300\times300$ grid points and a time step $dt=0.01$. The CLVs are computed just for $t=5$, so the phase \emph{T2}  and \emph{T5} includes just a few time steps. The forward and backward evolution is done in the time interval $T=[5,  \,\,10]$.
From $t=0$ to $t=5$ the algorithm passes through the initialization phase (see Appendix \ref{sec3:1-0}) to find the backward Lyapunov vectors bases. In Figure \ref{fig:13}
 it is shown the convergence for the spatial average of the scalar product between the starting random bases chosen. 

\subsection{Double gyre}
The parameters used for the double gyre example are  $A=0.1$, $\epsilon=0.1$ and $\omega=\pi/5$. The spatial domain  is $x=[0,\,\,2]$, $y=[0,\,\,1]$, with spatial resolution of $400\times200$ points and time step $dt=0.02$. 
Two experiments, \emph{DV1} and \emph{DV2}, involving different CLVs computational time interval are considered and  summarized in Tab. \ref{tab:1}.
\begin{table}
\begin{center}
\begin{tabular}{c |c| c| c| c| c}
    \hline
                      & \emph{T1} & \emph{T2} & \emph{T3} & \emph{T4} & \emph{T5} \bigstrut\\ \hline
    \emph{DV1} & $0\to5$ & $5\to10$ & $10\to15$ & $10\leftarrow15$ & $5\leftarrow10$\bigstrut\\ \hline
    \emph{DV2} & $0\to10$ & $10\to20$ & $20\to30$ & $20\leftarrow30$ &  $10\leftarrow20$\bigstrut\\
    \hline
    \end{tabular}
\end{center}
\caption{The five  temporal windows of the numerical algorithm used for the two experiments \emph{DV1} and \emph{DV2}. }
\label{tab:1}
\end{table}
During the initial phase of the numerical algorithm for the calculation of the CLVs, the average of the scalar product of the two initial random bases does not converge exactly to one (not shown here). This is because near the central regions of the two vortices of the double gyre, the expansion and contraction directions are not  well defined, the  CLVs become tangent to each other, and their separation is difficult to attain.

\subsection{Bickley jet}
The parameters used for this example are
$U=62.66$, $c_2=0.205U$, $c_3=0.461U$, $L_y=1.77\times10^6$, $\epsilon_2=0.04$, $\epsilon_3=0.3$, $L_x=6.371\pi\times10^6$, $k_n=2\pi n/L_x$.
The spatial domain considered is $x=[0,\,\,L_x]$, $y=[-2.25, \,\,2.25 ]L_y$, with a resolution of $500\times250$ grid points and a time step of $dt=1800$. 
Two different time windows evolutions are considered, experiments $BJ1$ and $BJ2$, and summarized in Tab. \ref{tab:2}.
\begin{table*}
\begin{center}
\begin{tabular}{c |c| c| c| c| c}
    \hline
                      & \emph{T1} & \emph{T2} & \emph{T3} & \emph{T4} & \emph{T5}\bigstrut \\ \hline
    \emph{BJ1} & $0\to1.89 $ & $1.89 \to  3.79$ & $3.79\to5.68$ & $3.79\leftarrow5.68$ & $1.89 \leftarrow3.79$\bigstrut\\ \hline
    \emph{BJ2} & $0\to3.79$ & $3.79\to7.57$ & $7.57\to11.36$ & $7.57\leftarrow11.36$ &  $3.79\leftarrow7.57$\bigstrut\\
    \hline
    \end{tabular}
\end{center}
\caption{The five  temporal windows of the  algorithm  used for the two experiments \emph{BJ1} and \emph{BJ2}. Times have been rescaled with $L_x / U$.}
\label{tab:2}
\end{table*}




\end{document}